\documentclass[12pt]{article}

\usepackage{scicite}
\usepackage{graphicx}
\usepackage{epsfig}
\usepackage[labelformat=empty]{caption}
\usepackage{amssymb}
\usepackage{amsmath}
\usepackage{mathrsfs}
\usepackage{color}
\usepackage{xcolor}
\usepackage{url}
\usepackage{threeparttable}
\usepackage{lipsum,pdflscape}

\newcommand\aj{{AJ}}%
\newcommand\apj{{ApJ}}%
\newcommand\apjl{{ApJ}}%
\newcommand\apjs{{ApJS}}%
\newcommand\aap{{A\&A}}%
\newcommand\araa{{ARA\&A}}%
\newcommand\pasp{{PASP}}%
\newcommand\mnras{{MNRAS}}%
 
\newcommand\nat{Nature} 
\newcommand\pasj{PASJ} 

\newcommand\figcaption{\def\@captype{figure}\caption}

\newcommand{\ergs}{${\rm erg \ cm^{-2} \ s^{-1}}$ }

\newcommand{\todo}{\ifmmode {\Huge \bullet} \else {\Huge$\bullet$}\fi}

\newcommand{\kms}{\ifmmode {\rm km\,s}^{-1} \else km\,s$^{-1}$ \fi}
\newcommand{\cc}{\hbox{cm$^{-3}$}}

\newcommand{\ergcms}{\ifmmode {\rm ergs\,cm}^{-2}\,{\rm s}^{-1} \else ergs\,cm$^{-2}$\,s$^{-1}$\fi}
\newcommand{\ergscmarc}{\ifmmode {\rm ergs\,cm}^{-2}\,{\rm s}^{-1} \else ergs\,s$^{-1}$\,cm$^{-2}$\,arcsec$^{-2}$\fi}
\newcommand{\ergcmsA}{\ifmmode{\rm ergs}\, {\rm cm}^{-2}\,{\rm s}^{-1}\,{\rm\AA}^{-1} \else ergs\, cm$^{-2}$\, s$^{-1}$\, \AA$^{-1}$\fi}
\newcommand{\ergcmsHz}{\ifmmode{\rm ergs\,cm}^{-2}\,{\rm s}^{-1}\,{\rm Hz}^{-1} \else ergs\,cm$^{-2}$\,s$^{-1}$\,Hz$^{-1}$\fi}
\newcommand{\phcms}{\ifmmode {\rm ph\,cm}^{-2}\,{\rm s}^{-1} \else ,ph\,cm$^{-2}$\,s$^{-1}$\fi}
\newcommand{\phcmsA}{\ifmmode {\rm ph\,cm}^{-2}\,{\rm s}^{-1}\,{\rm\AA}^{-1} \else ph\,cm$^{-2}$\,s$^{-1}$\,\AA$^{-1}$\fi}

\newcommand\Msun{\ifmmode M_{\odot} \else $M_{\odot}$\fi}
\newcommand\msun{\ifmmode M_{\odot} \else $M_{\odot}$\fi}
\newcommand\Lsun{\ifmmode L_{\odot} \else $L_{\odot}$\fi}
\newcommand\Zsun{\ifmmode Z_{\odot} \else $Z_{\odot}$\fi}
\newcommand{\mpyr}{$M_{\odot}{\rm yr}^{-1}$}

\newcommand{\Luv}{\ifmmode L_{1450} \else $L_{1450}$\fi}
\newcommand{\Lop}{\ifmmode L_{5100} \else $L_{5100}$\fi}
\newcommand{\Lthree}{\ifmmode L_{3000} \else $L_{3000}$\fi}
\newcommand{\lledd}{\ifmmode L/L_{\rm Edd} \else $L/L_{\rm Edd}$\fi}
\newcommand{\ledd}{\ifmmode L_{\rm Edd} \else $L_{\rm Edd}$\fi}
\newcommand{\lamLlam}{\ifmmode \lambda L_{\lambda} \else $\lambda L_{\lambda}$\fi}
\newcommand{\lbol} {\ifmmode L_{\rm bol} \else $L_{\rm bol}$\fi}
\newcommand{\llbol}{\ifmmode \log\left(\lbol/\ergs\right) \else $\log\left(\lbol/\ergs\right)$\fi}

\newcommand{\fuv}{\ifmmode f_{\lambda}\left(1450\AA\right) \else $f_{\lambda}\left(1450 {\rm \AA}\right)$\fi}
\newcommand{\fthree}{\ifmmode f_{\lambda}\left(3000\AA\right) \else $f_{\lambda}\left(3000{\rm \AA}\right)$\fi}
\newcommand{\fH}{\ifmmode f_{\lambda}\left(1.65\micron\right) \else
$f_{\lambda}\left(1.65\micron\right)$\fi}

\newcommand{\mbh}{\ifmmode M_{\rm BH} \else $M_{\rm BH}$\fi}
\newcommand{\lmbh}{\ifmmode \log\left(\mbh/\Msun\right) \else $\log\left(\mbh/\Msun\right)$\fi}

\newcommand \Hbeta {\ifmmode {\rm H}\beta \else H$\beta$\fi}
\newcommand \hb    {\ifmmode {\rm H}\beta \else H$\beta$\fi}
\newcommand  \mgii  {\ifmmode {\rm Mg}{\textsc{ii}} \else Mg\,{\sc ii}\fi}
\newcommand  \MGII  {\ifmmode {\rm Mg}\,{\sc ii}\,\lambda2798 \else Mg\,{\sc ii}\,$\lambda2798$\fi}
\newcommand  \siiv  {\ifmmode {\rm Si}\, {\sc iv}\ \else Si\,{\sc iv}\fi}
\newcommand  \SIIV  {\ifmmode {\rm Si}\,{\sc iv}\,\lambda1399 \else Si\,{\sc iv}\,$\lambda1399$\fi}
\newcommand  \civ  {\ifmmode {\rm C}\, {\sc IV}\ \else C\,{\sc IV}\fi}
\newcommand  \CIV  {\ifmmode {\rm C}\,{\sc iv}\,\lambda1549 \else C\,{\sc iv}\,$\lambda1549$\fi}
\newcommand  \NV  {\ifmmode {\rm N}\,{\sc v}\,\lambda1240 \else N\,{\sc v}\,$\lambda1240$\fi}
\newcommand  \nv  {\ifmmode {\rm N}\,{\sc v}\ \else N\,{\sc v}\fi}
\newcommand  \cv  {\ifmmode {\rm C}\,{\sc v}\ \else C\,{\sc v}\fi}
\newcommand  \LyA  {\ifmmode {\rm Ly}\,{\sc $\alpha$}\,\lambda1216 \else Ly\,{\sc $\alpha$}\,$\lambda1216$\fi}
\newcommand  \lya {\ifmmode {\rm Ly}\,{\sc $\alpha$}\ \else Ly\,{\sc $\alpha$}\fi}
\newcommand  \feii {\ifmmode {\rm Fe}\,{\textsc{ii}}\, \else Fe\,{\sc ii}\fi}

\newcommand  \aliii  {\ifmmode {\rm Al}{\textsc{iii}} \else Al\,{\sc iii}\fi}

\newcommand  \CIII  {\ifmmode {\rm C}\,{\sc iii]}\,\lambda1909 \else C\,{\sc iii]}\,$\lambda1909$\fi}
\newcommand  \oi    {\ifmmode \left[{\rm O}\,{\textsc i}\right] \else [O\,{\sc i}]\fi}
\newcommand  \OI    {\ifmmode \left[{\rm O}\,{\textsc i}\right]\,\lambda6300 \else [O\,{\sc i}]$\,\lambda6300$ \fi}

\newcommand  \oii   {\ifmmode \left[{\rm O}\,{\textsc ii}\right] \else [O\,{\sc ii}]\fi}
\newcommand  \OII   {\ifmmode \left[{\rm O}\,{\textsc ii}\right]\,\lambda3727 \else [O\,{\sc ii}]\,$\lambda3727$ \fi}
\newcommand  \oiii  {\ifmmode \left[{\rm O}\,{\textsc iii}\right] \else [O\,{\sc iii}]\fi}
\newcommand  \OIII  {\ifmmode \left[{\rm O}\,{\textsc iii}\right]\,\lambda5007 \else [O\,{\sc iii}]\,$\lambda5007$\fi}

\newcommand{\lmg}{\ifmmode L\left(\mgii\right) \else $L\left(\mgii\right)$\fi}
\newcommand{\fwmg}{\ifmmode {\rm FWHM}\left(\mgii\right) \else FWHM(\mgii)\fi}
\newcommand{\fwciv}{\ifmmode {\rm FWHM}\left(\civ\right) \else FWHM(\civ)\fi}
\newcommand{\fwhm}{\ifmmode {\rm FWHM} \else FWHM\fi}

\def\ergs{\rm erg~s^{-1}}

\def\civ{C {\sc iv}}
\def\feii{Fe {\sc ii}}
\def\mgii{Mg~{\sc ii}}
\def\oiii{[O~{\sc iii}]}

\usepackage{times}

\newenvironment{sciabstract}{%
\begin{quote} \bf}
{\end{quote}}

\newcounter{lastnote}

\title{Discovery of spectacular quasar-driven superbubbles in red quasars}

\author
{Lu Shen$^{1,2,3}$, Guilin Liu$^{1,2\ast}$, Zhicheng He$^{1,2\ast}$, Nadia L. Zakamska$^{4\ast}$, \\Eilat Glikman$^{5}$, Jenny E. Greene$^{6}$, Weida Hu$^{1,2,7}$, Guobin Mou$^{8}$, \\Dominika Wylezalek$^{9}$,  David S. N. Rupke$^{10}$\\
\\
\normalsize{$^{1}$CAS Key Laboratory for Research in Galaxies and Cosmology, Department of Astronomy, }\\
\normalsize{University of Science and Technology of China, Hefei, Anhui 230026, China}\\
\normalsize{$^{2}$School of Astronomy and Space Science, University of Science and Technology of China, }\\
\normalsize{Hefei 230026, China}\\
\normalsize{$^{3}$Department of Physics and Astronomy, Texas A\&M University, }\\
\normalsize{College Station, TX 77843-4242 USA}\\
\normalsize{$^{4}$Department of Physics \& Astronomy, The Johns Hopkins University, }\\
\normalsize{3400 North Charles Street, Baltimore, MD 21218, USA}\\
\normalsize{$^{5}$Department of Physics, Middlebury College, Middlebury, VT 05753, USA}\\
\normalsize{$^{6}$Department of Astrophysical Sciences, Princeton University, Princeton, NJ 08544, USA}\\
\normalsize{$^{7}$Department of Physics, University of California, Santa Barbara, } \\
\normalsize{Santa Barbara, CA 93106, USA}\\
\normalsize{$^{8}$Department of Astronomy, School of Physics and Technology, Wuhan University,} \\
\normalsize{Wuhan 430072, China}\\
\normalsize{$^{9}$Zentrum f\"{u}r Astronomie der Universit\"{a}t Heidelberg, Astronomisches Rechen-Institut, }\\
\normalsize{M\"{o}nchhofstr 12-14, D-69120 Heidelberg, Germany}\\
\normalsize{$^{10}$Department of Physics, Rhodes College, Memphis, TN 38112, USA}\\
\\
\normalsize{$^\ast$To whom correspondence should be addressed; E-mail:  glliu@ustc.edu.cn, }\\
\normalsize{zcho@ustc.edu.cn, zakamska@jhu.edu.}
}

\date{}

\begin{document} 


\baselineskip24pt


\maketitle


\begin{sciabstract}
{Quasar-driven outflows on galactic scales are a routinely invoked ingredient for galaxy formation models. }
We report the discovery of ionized gas nebulae 
surrounding three luminous red quasars at $z{\sim}0.4$ from Gemini Integral Field Unit (IFU) observations. 
All these nebulae feature unprecedented pairs of ``superbubbles'' {extending $\sim$20} kpc in diameter, and the line-of-sight velocity difference between the red- and blue-shifted bubbles reaches up to $\sim$1200 km~s$^{-1}$. Their spectacular dual-bubble morphology (in analogy to the Galactic ``Fermi bubbles'') {and their kinematics provide unambiguous evidence for galaxy-wide quasar-driven outflows, in parallel with the quasi-spherical outflows similar in size from luminous Type-1 and -2 quasars at concordant redshift. }
These bubble pairs manifest themselves as a signpost of the {short-lived} superbubble ``break-out'' phase, when the quasar wind drives the bubbles to escape the confinement from the dense environment and plunge into the galactic halo with a high-velocity expansion. 
\end{sciabstract}



\section*{Introduction}

In the evolutionary paradigm of feedback from active galactic nuclei (AGN), the merger-driven accretion power in quasars succeeds in driving a powerful wind{. This wind is capable of rapidly sweeping away cold gas and dust, clearing the future fuel of galaxy} for star formation, and {turning} the system into an unobscured quasar {\cite{Tabor1993, scannapieco2004, DiMatteo2005, Hopkins2010}}. As the most luminous quasars at every epoch, red quasars are a natural place to hunt for energetic outflows, and their morphology that strongly points to a connection with merging hints for important clues to quasar/galaxy evolution \cite{Urrutia2008, Glikman2015, Wylezalek2022}. 

{In this paper, we present 3 red quasars at $z \sim 0.4$ with unambiguous signatures of superbubble pairs. We conduct IFU mapping of these targets utilizing the Gemini-North Multi-Object Spectrographs (GMOS-N) \cite{AllingtonSmith2002} equipped on the Gemini 8m Telescope. The identical} setup and analysis strategy have been successfully used to investigate the ionized gas nebulae surrounding 11 Type-2 and 12 Type-1 highly luminous radio-quiet quasars at z $\sim$ 0.5 \cite{Liu2013a, Liu2013b, Liu2014}. These red quasars are selected from the FIRST-2MASS (F2M) sample by cross-matching the FIRST survey \cite{Becker1995}, Two Micron All Sky Survey (2MASS\cite{Skrutskie2006}) and the Guide Star Catalog II (GSC-II) with selection criteria of $J - K >$ 1.7 and $R - K >$ 4.0  \cite{Glikman2007, Glikman2012}. 
For comparison purposes, we further require them to have a WISE luminosity matched to our previously observed Type 1 and 2 quasars ($\lambda L_\lambda[12\mu m] \sim 10^{45-45.9}$ erg s$^{-1}$, see Methods and Table S2), {to be non-radio-loud (see calculation on radio loudness in Methods and Table S2), and to situate at similar redshift ($z\sim0.4$).}  


\section*{Results}

{We observe all of our targets with GMOS-N IFU in i band (7050–8500 \AA) so as to cover the rest-frame wavelength range 4086--5862 \AA\ that encloses the \oiii\ $\lambda5007$ \AA\ emission line. We perform data reduction using the standard Gemini package for IRAF and produce the final science data cubes with a spaxel scale of $0.\!\!^{\prime\prime}05$. We further perform flux calibration, Point Spread Function (PSF) subtraction, and a multi-Gaussian fit to the profile of the \oiii\ $\lambda$5007\AA\ emission line in each spaxel (See Methods for details). We characterize the morphology and kinematics of these ionized gas nebulae by mapping the spaxel-by-spaxel distribution of three parameters, following L13b: the integrated \oiii\ emission (denoted $Int$) to characterize the \oiii\ surface brightness, the median velocity ($V_{\rm med}$, the 50\%-th quantile) to characterize the line-of-sight velocity, and the velocity interval that encloses 80\% of the total \oiii\ flux ($W_{80}$, i.e. the difference between the 10\%-th and 90\%-th quantiles) to characterize the velocity dispersion. }

All three red quasars show extended ionized gas nebulae, with each of them having a pair of red-shifted and blue-shifted {bubbles.} 
{In Fig. 1, Fig. 2 and Fig. 3, we map the distribution of the above-mentioned parameters across the entire quasars and their individual blue- and red-shifted bubbles. }
{We find clear signatures of superbubbles in their well-resolved dual-bubble morphology, which is further aided by the information rendered by \oiii\ kinematics:} the narrowness of the \oiii\ line in the outskirts of the bubbles is in line with a geometrically thin shell, and the well-defined double-peak velocity profiles near the galactic centers indicate where the pair of bubbles overlap {in projection}. These velocity profiles facilitate separation of the emission from the blue- and red-shifted bubbles through meticulous \oiii\ line-fitting, so as to track down each individual bubble in a remarkably clean manner {(also see Methods and Fig. S4, S5, S6).} 

{The existence of superbubble pairs provides} unambiguous evidence for the existence of quasar-driven outflows{, in analogy to the Fermi bubbles seen in the Milky Way \cite{Su2010, Predehl2020}. }
{The fact that the \oiii\ peaks and the velocity gradient lie along the same axes also yields a natural outflow interpretation, as these \oiii\ bubbles are clearly photoionized by the quasar \cite{Harrison2014, Vayner2021}. } 
The high-velocity differences and the superposition of red-shifted and blue-shifted bubbles further support an outflow origin and exclude alternative possibilities such as an inflow \cite{Dekel2009, Steidel2010} or a rotating galaxy \cite{Reyes2011, Pelliccia2017, VegaBeltran2001}. 

{We compare the morphology and kinematic properties of these ionized gas nebulae with those surrounding type 1 and type 2 quasars. }
{To do this, we measure the outflow sizes using the spaxels with $>5\sigma$ \oiii\ detection. As reported in Table 1, this directly observed radius ($R_{5\sigma}$) ranges from 8.7 to 12.3 kpc}. We also report the ``intrinsic'' isophotal radius ($R_{\rm int}$) and {ellipticity ($\epsilon_{\rm int}$) of the best-fit
ellipse} measured at a surface brightness of $10^{-15}/(1 + z)^{4}$ {erg s$^{-1}$ cm$^{-2}$ arcsec$^{-2}$} (so that the cosmological dimming effect is corrected, following Liu13a). We find $R_{\rm int}$ = {8.6 -- 11.7} kpc, {similar to} that of type 1 ($\langle R^{T1}_{int} \rangle$ = 10.7$\pm$1.7 {kpc}) and type 2 quasars ($\langle R^{T2}_{int}\rangle$ = 12.9 $\pm$3.4 {kpc}).  
The median ellipticity of these dual-bubble red quasar nebulae is $\epsilon_{\rm int} \sim 0.75$, in contrast to the quasi-spherical type 1 and type 2 quasar nebulae with a median of $\epsilon^{T1}_{int} =0.12$ and $\epsilon^{T2}_{int} =0.18$, respectively.

Spatially resolved and highly organized velocity structure is present in every nebula (see Fig. 1, Fig. 2 and Fig. 3),
and blue- and red-shifted emission predominantly reside on the opposite sides of the quasar. 
The maximum projected velocity difference between the red-shifted and the blue-shifted regions is $\delta v_{max}$ = 394--1216 km s$^{-1}$, {substantially} higher than that of type 1 and 2 quasars found in L13b: 83-576 km s$^{-1}$ and L14: 89-522 km s$^{-1}$,
{implying for energetic outflows from these red quasars. }

The median and the maxima of $W_{80}$ are comparable to that of type 1 and type 2 quasars (up to $\sim$1000 km s$^{-1}$). An evident belt-like feature is commonly seen in the $W_{80}$ maps across the center and perpendicular to the direction of the bubble expansion, where velocity dispersion {is} higher than the rest of the nebula. This feature is mainly due to the superposition of red- and blue-shifted bubbles in projection (as already mentioned above), while the actual  {line width of} \oiii\ in each individual bubble remains consistently narrow throughout its entire extent (see Methods and Fig. S4, S5, S6; this phenomenon is also seen in our simulation results ({Fig. 4}).

\section*{Simulation}

Previous theoretical\cite{scannapieco2004,di2005, murray2005, ciotti2009,hopkins2009,ostriker2010} and observational \cite{arav2008,moe2009,arav2013,arav2018,hamann2011,hamann2019,he2019, Chen2022} {works suggest that high velocity ($\gtrsim 10^3$ km s$^{-1}$) quasar winds may play a critical} role in the evolution of their host galaxies. 
{Numerical simulations are an indispensable tool in understanding the effects induced by outflows, which can model and predict nonlinear and complex processes prohibitable to analytical analysis. However, previous simulations have been primarily focused on AGN-driven jets \cite{Wagner2012, Tanner2022}, ultra-fast outflows with velocity in the range of $10^3$ - $10^4$ km s$^{-1}$ \cite{Wagner2013}, starburst-driven outflows \cite{Schneider2018}, and more general galactic outflows driven by supernovae and supermassive black holes in the cosmological simulation TNG50 \cite{Nelson2019}.} 
{Simulations providing a direct link to observational data remain scarce, though significant progress has been made to combine a solid physical basis and predictability of observations \cite{Yuan2018}. 
In order to determine the outflow parameters in a manner more accurate than conventional analytical calculations based on oversimplified assumptions, and to} help constrain the conditions under which the formation of the observed superbubbles becomes realistic, we conduct a data-oriented two-dimensional (2D) hydrodynamics (HD) simulation. 
{We emphasize that our simulation is limited to being an auxiliary tool for the above purposes, in contrast to those deriving from fundamental physical principles.}

{In our simulation, we assume} a typical quasar with a back hole (BH) mass of $\mbh =10^9\Msun$ and an Eddington ratio of 0.3, corresponding to a bolometric luminosity of $\lbol =3.9\times 10^{46}\ergs$, a value matched to those of our red quasars. 
{We initiate the outflowing motion by artificially placing a nozzle in the galactic center. The initial kinetic energy (at the nozzle) is chosen based on the observed power of BAL outflows found in the range of 1--10\% percent of their \ledd\ \cite{moe2009, borguet2013, chamberlain2015, he2019, Choi2020, He2022}.} The half-opening angle of the wind and the ionization cones are assumed to be the same ($\Phi=45^{\circ}$). {The radiation transfer models in the simulation are adopted following ref.\cite{proga2000}.
We then employ the software \textsc{Cloudy}\cite{Ferland2017} to obtain the \oiii\ emission coefficient based on the gas density and ionization parameters from the simulation. Finally, a 3D system is obtained by rotating the 2D simulation around the y-axis and re-projecting the 3D structure to 2D with an assigned inclination angle} (see {Methods} for the details of the simulation setup, initial conditions, and calculations). 

To reproduce the observed superbubbles, we {find it feasible to maintain the initial power of the quasar wind} to be a constant $P_{\rm wind} = 3.35\times 10^{45}\;\ergs=8.6\%\; \lbol = 2.6\%$ of Eddington luminosity (\ledd) lasting over the entire simulation. 
In {Fig. 4}, we present the resultant simulated \oiii\ maps and the corresponding emission line profiles after having the simulation evolve for 12 Myr, where an inclination angle of $i$=30$^{\circ}$ is adopted. As seen in the figure, the simulated bubble pair is consistent with our observations, where the peanut-like morphology, the well-organized velocity structures and the central belt-like high $W_{80}$ regions are all reproduced. 

As an experiment, we render the simulated system to evolve for an additional period of 8 Myr, finding that the simulated nebula gradually grows into a quasi-sphere (see Methods and Fig. S7). This result implies that the lifetime of the observed superbubbles may be as brief as $\sim10$ Myr, roughly in line with the small space density of red quasars $\sim$15-30\% in the total radio-selected quasars \cite{Glikman2012, Glikman2018, Glikman2022}. 
Hence, our simulation reveals that the formation of such superbubbles requires a continuous energetic wind lasting over $\sim$12 Myr with a kinetic luminosity of 2.6\% of the quasar's Eddington luminosity, and the bubble pair quickly grows into a quasi-spherical morphology. If the quasar is reasonably assumed to operate with a 10\% duty cycle, then the kinetic-to-Eddington luminosity ratio has to scale up by a factor of 10, placing these outflows among the most powerful ones reported hitherto \cite{Arav2020, Capellupo2014}. 
{We note that the simulated bubbles do not depend on the lifetime of an individual episode of quasar activity, which is found in the range of $\sim 10^3 - 10^5$ years \cite{Martini2003, Shen2021}. This is because, as seen in our simulation, the bubbles, once produced, disappear slowly over a much longer timescale. As a result, after the first inflated bubble clears the space inside it, the bubble structure remains in place, until the quasar turns on the next time, then the engulfed material quickly reaches the shell of the previous bubble and continues to push it. }

Alternatively, a quasi-spherical nebula can be obtained by increasing the opening angle of the wind/ionization cone (see Fig. S7), rendering this angle a sensitive parameter to shape the morphology of the system. Therefore, our simulation raises the possibility that the opening angle of red quasars is smaller than that of type 1 and 2 quasars (Liu13a, Liu13b, Liu14) when they are set to evolve for the same period.  

{Even though} our simplistic simulation is incapable of disentangling certain evolutionary and geometric effects, the combination of observation and simulation is beneficial to the {understanding of the} formation of the spectacular bubble pairs discovered in this work. Our simulation is consistent with the scenario that when the bubbles have escaped from the confinement of a high-density environment (e.g. a galactic disc), {the wind dives rapidly into the galactic halo. } 
Therefore, we conclude that the red quasars reported in this work are caught in a {short-lived} supper-bubble ``break-out'' phase on their multi-stage evolutionary track.

\section*{Discussion}


{All three red quasars show a pair of \oiii\ superbubbles with a projected spatial extent of {$\sim$20 kpc} in diameter and a highly organized velocity structure with a line-of-sight velocity difference of $\sim$300--1200 km~s$^{-1}$. } 
{We conduct a data-oriented 2D hydrodynamics simulation, revealing that the formation of such super-bubbles requires an energetic wind predicted to be observable in a brief time scale ($\sim$10 Myr). }

{The spectacular bubble morphology provides unambiguous evidence for quasar-driven outflows. The existence of bubbles also offers unique opportunities to measure the energy and momentum, which may be inaccessible for winds in other morphologies \cite{Greene2012, Nesvadba2006}. 
Our superbubbles is morphologically similar to the Fermi bubbles in the Milky Way \cite{Su2010, Predehl2020} and the superbubbles seen in local starburst galaxies \cite{Sakamoto2006, Tsai2009}. 
Superbubbles have been observed in a handful of low-redshift type-2 quasars: J1356+1026 at $z=0.123$ with extended X-ray emission (20 kpc) co-spatial with the ionized gas \cite{Greene2014}, and J1430+1339 at $z=0.085$ (known as the ``Teacup AGN'') with a pair of $\sim$10 kpc radio bubbles \cite{Harrison2015}, \oiii\ emission bubbles \cite{Keel2015}, and an arc of X-ray emission tracing the radio and ionized gas emission \cite{Lansbury2018}. 
Recently, a one-sided superbubble is reported to be driven by a non-broad absorption line (non-BAL) quasar at $z=0.631$, HE 0238-1904, and the associated \oiii\ emission reaches a projected distance of $\sim$55 kpc \cite{Zhao2023}. In addition to the bubbles themselves, these {works} suggest that quasar-driven superbubbles can imprint themselves in the X-ray and radio wavelengths when blowing their shells, {so that important physical parameters (e.g. temperature and column density of ionized gas) are measurable. Along with future X-ray and radio observations, we expect that the Integral Field Spectrograph (IFS) to be equipped on the Chinese Space Station Telescope (CSST) with a higher spatial resolution of $\sim0.^{\prime\prime}2$ will allow for in-depth scrutinization of the kinematics of these nebulae and superbubbles.}

Such dusty energetic winds and associated shocks are also expected to be responsible for generating radio emission in red quasars \cite{Zakamska2014, CalistroRivera2021, Klindt2019, Smith2020, Glikman2022}. The radio-quiet/intermediate nature of our red quasars (with radio loudness in the range of $R \sim -4.9$--4.2; see Table S2 and Methods) supports the scenario that their radio emission may originate, at least partly, from a dusty shocked wind. 
{Our three sample red quasars have a radio power of $L_{\rm1.4GHz} \sim 10^{24.6 - 25.0}$ W~Hz$^{-1}$ as per their peak flux densities at 1.4 GHz as measured in FIRST (see Table S2). F2M1618 shows a higher radio power ($L_{\rm 1.4GHz} \sim 10^{25.4}$ W~Hz$^{-1}$ according to the flux density measured in a larger beam size, which indicates extended radio emission. }
{The current radio data do not allow us to conclude whether the radiatively driven outflows cause shocks that result in synchrotron emission, or potientially existent weak radio jets {are} driving the outflows, in view of the suggestion that radio jets can play a role in driving outflows even in radio-quiet AGNs \cite{Morganti2015, Wylezalek2018}.}
Future spatially-resolved radio maps, possibly from JVLA observations, promise to help pin down the origin of the radio emission.


One of our red quasars, F2M0830, shows evidence of an ongoing merger \cite{Urrutia2008}. \cite{Urrutia2008} find {that} 12 out of 13 red quasars {show} 
recent or ongoing interaction, suggesting that major, gas-rich  mergers may be the origin of quasar activities, in line with \cite{Hopkins2008, Treister2012}. 
{However, the time scales of merging and outflows are substantially different. Galaxy merging can last for 1--3 Gyr years, and be observable for $\sim$0.2--2 Gyr, depending on the method used to identify the merger, the gas fraction, separation of the merging galaxies, and their relative orientation \cite{Lotz2008, Lotz2010}. However, based on our simulation, we estimate the observable time-scale of the bubble morphology to be $\sim$10 Myr, substantially briefer than that of a merging event. } 



{Another intriguing fact seen in our \oiii\ surface brightness maps is that the red-shifted bubbles are, in general, more luminous in \oiii\ than their blue-shifted counterparts by $\sim$0.15 dex (Table 1).} 
{This is unlikely caused by the dust torus surrounding the central engine, or the galaxy-wide dusty disk of the host galaxy, in which case higher extinction in the red-shifted bubble is expected, instead. } 
{On the contrary, we suspect that dust is not uniformly distributed within an individual cloud, but is preferably distributed on the side of the cloud further away from the galaxy center, so that the flux of the blue-shifted component is suppressed. This phenomenological explanation is schematically illustrated in {Fig. 5}. } {The higher dustiness at the far side away from the radiation source is potentially due to dust remnants from mergers, or dust distributed along the polar axis extending out to a few hundred parsecs likely associated with dusty outflows \cite{GarciaBurillo2021}. 
Unfortunately, the host galaxies are minimally detected in our IFU data.} We expect that \textit{JWST} mapping in the mid-infrared or campaigns conducted at sub-millimeter {wavelengths may reveal the dust distribution in these host galaxies, providing further constraints on this interpretation.} 

{It has been predicted theoretically that an outflow with a kinetic luminosity of $\sim$0.5\%-5\% of the AGN’s bolometric luminosity can act as an agent of significant feedback \cite{Hopkins2010}. Our simulation reveals that the formation of such superbubbles requires kinetic luminosity of 2.6\% of the Eddington luminosity, corresponding to 8.6\% of the bolometric luminosity, implying for effective feedback at work. } {In fact, our simulation renders a mass outflow rate of the final quasar wind one order of magnitude higher than that of the initial wind from the nozzle, implying a highly efficient interaction with the ISM that promises to shape the evolution of its host galaxy. }
{It has been suggested that the fast winds from quasars may impact structures within the interstellar medium and deposit energy into the intergalactic medium \cite{scannapieco2004}. }
Future {campaigns to be conducted at sub-millimeter wavelengths employing} ALMA or NOEMA may place {improved} constraints on the impact of winds on cold gas distribution, and {consequentially} on quenching {and/or triggering of} the star formation activity {in} the host galaxy.

\section*{Materials and Methods}
\smallskip
\noindent\textbf{IFU observation and data analysis}
\smallskip

In this section, we describe the IFU observations and the processes in IFU data analysis to obtain the maps of line-integrated surface brightness, line-of-sight velocity, and the velocity dispersion of the \oiii\ emission line. 


{In our IFU campaign, we adopt the two-slit mode with a 5$^{\prime\prime}\times$ 7$^{\prime\prime}$ field of view (corresponding to $\sim$28$\times$39 kpc$^2$ for our quasars at $z \sim 0.4$). The science field of view is sampled by 1000 contiguous $0.\!\!^{\prime\prime}2$ diameter hexagonal lenslets, and simultaneous sky observations are obtained by 500 lenslets located $\sim1^\prime$ away. The seeing at the time of our observations is $\sim0.\!\!^{\prime\prime}4$ (2.1 kpc at $z=0.4$), as determined by measuring the full-width-half-maximum (FWHM) of the profile of multiple field stars in the acquisition image taken right before the science exposure using \textsc{psfex} \cite{Bertin2011}. 
All of the targets are observed in i band (7050–8500 \AA) so as to cover the rest-frame wavelength range 4086--5862 \AA\ that encloses the \oiii\ emission line. 
To ensure that none of the important emission lines in this region is severely hindered by the slit gaps, we tune the central wavelength to either 760 or 800 nm, according to their respective redshifts. 
The employed R400-G5305 grating has a spectral resolution of R = 1918. At the wavelengths of \oiii\ for these three red quasars, this corresponds to a full width at half maximum (FWHM) of $\sim$167 km s$^{-1}$ \cite{Liu2013b}, narrower than our observed \oiii\ line in all cases, rendering the velocity profiles spectrally resolved. For each object, we take two exposures of 1620 sec per each without a spatial offset. Relevant information on our Gemini-GMOS observations is summarized in Table S1. 
We perform the data reduction using the Gemini package within IRAF version 1.14. 
The spaxel scale of the final science data cubes is set to be $0.^{\prime\prime}05$. }



We flux-calibrate our data using the spectra from the Extended Baryon Oscillation Spectroscopic Survey (eBOSS) for F2M1106 and F2M0830 \cite{Dawson2016}, and the spectra obtained at the W. M. Keck Observatory with the Echellette Spectrograph and Imager \cite{Sheinis2002} for F2M0830 \cite{Glikman2007, Glikman2012}. {We follow } the method presented in ref.\cite{Liu2014}. 
In short, we scale the IFU data against the eBOSS and ESI spectra by mimicking the observing conditions of these existing spectra. 
The eBOSS spectra of F2M1106 and F2M0830 are collected by fibres with a 2$^{\prime\prime}$ diameter at a median FWHM$_{\mathrm{eBOSS}}$ of $\sim1.5^{\prime\prime}$ and $\sim1.2^{\prime\prime}$, respectively. 
In order to mimic the observing condition of eBOSS spectra, the IFU image at each wavelength is convolved with a Gaussian kernel with an FWHM of $\sqrt{\mathrm{FWHM_{eBOSS}^2} - \mathrm{seeing}^2}$, where the seeing of IFU data is listed in Table S1. The GMOS spectra between the rest-frame 4980 and 5200 \AA\ are extracted using a 2$^{\prime\prime}$-diameter circular aperture. 
The ESI spectrum of F2M0830 is taken with an $20^{\prime\prime} \times 1^{\prime\prime} $ slit, and placed at an angle of 29$^\circ$ from North to East. The GMOS spectrum is extracted using a $4^{\prime\prime} \times 1^{\prime\prime}$ box with the same angle centered on the maximum of the integrated flux of the IFU data cube. The length of this box is smaller than that of ESI due to the field of view of the GMOS IFU. Thus, we compare the resultant spectra to the ESI spectra in the rest-frame wavelength range of 5030--5200 \AA, as the continuum flux is dominated by the quasar, which is not affected by the size of the box.


Quasar light scattered by the interstellar matter\cite{Zakamska2006}, star formation in the quasar host\cite{Letawe2007, Silverman2009} and the PSF itself might all contribute to the continuum emission. 
To focus on the kinematics of nebulae gas, we construct a PSF modeled for each target and subtracted it by scaling to its quasar spectrum from the IFU data cube. 
The PSF is constructed by interpolating between median images in the two rest-frame wavelength intervals of 4970\AA--4980\AA\ and 5030\AA--5050\AA\ and normalizing the peak flux of each image to unity. The wavelength intervals are chosen free of \oiii\ and \Hbeta\ line emission. 
Due to the wavelength coverage of the IFU data of F2M1106 and F2M0830, their PSFs are constructed as the normalized median image in a single wavelength interval of 5030\AA--5050\AA. The PSFs of the three red quasars are shown in Fig. S1. 

The quasar spectrum is constructed by removing the recovered red- and blue-shifted [O \textsc{iii}] component in the spectra of the central spaxels. The central spaxel is determined by the maximum of the integrated flux between rest-frame 4900 and 5100 \AA. We then reconstruct the combined spectra of the central 5$\times$5 spaxels by fitting a three/four Gaussian model, after removing Fe \textsc{ii} emission. 
The \oiii\ doublet lines are fitted with the same central velocity and velocity dispersion. }
The combined and fitted spectra are shown in Fig. S2. 
The Fe \textsc{ii} emission is determined by fitting to the continua in the rest-frame 5100--5250 \AA\ using the Fe \textsc{ii} template from \cite{Boroson1992} and smoothed using a Gaussian kernel, whose width is one of the free fitting parameters.
We find Fe \textsc{ii} emission to be substantial in the central spectrum of F2M1106, as shown in Fig. S2, but undetected in the other two red quasars. 
To confirm that the Fe \textsc{ii} emission of F2M1106 is dominated by the quasar, we perform the same procedure on each spaxel. The intensity map of Fe \textsc{ii} emission, as shown in Fig. S3, reveal a point source with the same size as the PSF and centered at the central spaxel. Therefore, Fe \textsc{ii} is removed during PSF subtraction. 
The central spectrum of F2M1106 is not well fitted at $<4980$\AA\, possibly due to additional Fe and/or the broad wing of the H$\beta$ line. 


After flux calibration and subtraction of the PSF from the IFU data cubes, we perform a multi-Gaussian fit to the \oiii\ $\lambda$5007\AA\ line profile, so that a noiseless model of the line is obtained in every spatial pixel\cite{Liu2013b}. 
As described in L13b, up to 3 Gaussian components are needed for these fits. The actual number of employed Gaussians is determined by comparing the reduced $\chi^2$ values as a function of the number of components. 
The uncertainty of the spectrum of each spaxel is the standard deviation of spectra in regions with \oiii $< 3\sigma$, which is used to calculate the reduced $\chi^2$ values. 
We then compute the \oiii\ line {properties (i.e., intensity, median velocity, velocity dispersion)} in every spatial position from the multi-Gaussian fit (instead of the observed profile). 
In addition, we subtract additional continuum contribution using spectra in $\lambda_{rest}\sim$ 5050-5100\AA. Only a minimal continuum residual is left after the PSF subtraction due to the spaxel variation of IFU data. 
Individual Gaussian is then assigned to red- and blue-shifted bubbles according to their mean wavelength. 
Following ref.\cite{Whittle1985} and L13b, we measure two quantities that characterize the line-of-sight velocity and velocity dispersion of the ionized gas: the median velocity ($v_{\rm med}$) that bisects the total area underneath the \oiii\ emission line profile, and the velocity interval that encloses 80\% of the total \oiii\ emission centered at the median velocity ($W_{80}$). As noted in previous references, $W_{80}$ is more sensitive to the weak broad wings of a non-Gaussian profile, but is similar to FWHM for a Gaussian profile ($W_{80}=1.088\times {\rm FWHM}$). 
Examples of spectra in the wavelength vicinity of \oiii\ in a number of representative spaxels are shown in {Fig. S4, Fig. S5, Fig. S6}. 
These spectra are selected across the entire \oiii\ outflow to demonstrate the complexity of the \oiii\ line profile. 



\smallskip
\noindent\textbf{Bolometric luminosity of red quasars} 
\smallskip

We estimate the bolometric luminosity of our red quasars to guide our simulation. 
This task is nontrivial, as red quasars are dust-obscured at wavelengths from X-ray to optical and maybe even mid-infrared \cite{Zakamska2008, Kim2018}. 
First, we adopt the bolometric luminosity estimated from luminosity at rest-frame 12$\mu$m extrapolated from WISE photometry, rendering a bolometric correction factor of 9 \cite{Richards2006} (see Table S2). The results are in the range of $10^{46.5 - 46.8}$ erg s$^{-1}$. 
However, these bolometric luminosities may be either underestimated, as the adopted standard bolometric correction is derived from unobscured quasars, or overestimated, if other sources (e.g. dust in the host) contribute to the rest-frame 12$\mu$m flux as well. 

An alternative approach is using the luminosity at rest-frame 5100 \AA\ and applying a bolometric correction of BC$_{5100}$ = 9.65 \cite{Shen2011, Richards2006}, along with additional dust extinction correction. 
We adopt the extinction $E(B-V)$ from ref. \cite{Glikman2012} by fitting a reddened quasar continuum to a full spectrum. The intrinsic shape of quasar continua is assumed to be a Gaussian distribution with $f\nu \propto \nu^{\alpha}$ (See ref. \cite{Glikman2012} for details). 
This approach leads to resultant bolometric luminosities in the range of $10^{46.0 - 46.3}$ erg s$^{-1}$. 
Hence, we obtain results {consistent within a factor of 3-5} from these two methods.

\smallskip
\noindent\textbf{Radio-loudness of red quasars} 
\smallskip

Radio-loudness is defined as the ratio of radio to optical emission with radio-loud objects typically possessing powerful collimated radio jets. However, the presence of reddening and extinction at optical wavelengths may render red quasars to be artificially misclassified as radio-loud. 
Therefore, we adopt the radio-loudness definition following \cite{Glikman2022, Klindt2019}:
\begin{equation*}
    R = \mathrm{log_{10}}\left(\frac{1.4\times10^{16} L_{\mathrm{1.4GHz}}}{L_{\rm 6\mu m}}\right),
\end{equation*} 
where $L_\mathrm{{1.4GHz}}$ is in units of W Hz$^{-1}$ and $L_{\rm 6\mu m}$ in erg/s. 
The latter is less sensitive to dust extinction but it still probes the quasar continuum. 
Radio-quiet objects are defined to have $R<-4.6$, radio-intermediate $R = -3.5~\sim~-4.6$, and radio-loud $R>$ -3.5.  
Under this definition, F2M0830 and F2M1106 are radio-quiet, while F2M1618 is radio-intermediate, regardless of radio fluxes from FIRST or from ref. \cite{Glikman2007} in an NVSS-like beam size. 
All of them are point-like sources in the FIRST images. Only F2M1618 has a major axis larger than the $\sim5^{\prime\prime}$ beam size of FIRST. 
Hence, our red quasars are radio-selected objects, but are in the radio-quiet/intermediate regime. Their radio emission may originate, at least partly, from shocked winds \cite{Zakamska2014, Klindt2019, Glikman2022}.

\smallskip
\noindent\textbf{Simulation setup}
\smallskip



We conduct a two-dimensional (2D) hydrodynamics (HD) simulation using \textsc{zeusmp}\cite{Stone1992, Hayes2006} code to reproduce the observed quasar outflow, in order to fully understand its physical conditions. In this section, we describe in detail the simulation setup and initial conditions.

The simulation is performed in a two-dimensional (2D) polar coordinate system. There are 1750 grid cells in the radial (r) equally spaced in logarithm from r = 0.5 kpc to r = 50 kpc and 600 grid cells equally spaced in the azimuthal from 0 along the Y-axis to $\pi/2$ rotate clockwise. The logarithm radial grid provides high resolution to capture the injection of wind in the simulation grid. 
The azimuthal range is chosen under the assumption of symmetrical wind/bubble, thus, the results in the other three quadrants are the reflection of that in the first quadrants.

We adopt a BH mass of $\mbh =10^9\ \Msun$, corresponding to an Eddington luminosity of $\ledd \sim 1.3\times 10^{47}\ \ergs$, and a typical quasar with Eddington ratio of 0.3, corresponding to a bolometric luminosity of $\lbol = 3.9\times 10^{46}\ \ergs$, which is consistent with those of our red quasars ($\nu L_{bol, 12\mu m} =  3.2 - 6.5 \times 10^{46}\ \ergs$). 

The initial environment where the wind is propagating includes a spherical central dense nebular and a spherical surrounded diffuse ISM. 
The dense nebular has an average number density of $10^2\ \cc$ within a radius of 2 kpc. 
Following the $M-\sigma$ relation \cite{Kormendy2013} and accounting for the stellar rotational velocity that remains constant with increasing distance away from the galactic center, we use a constant velocity dispersion $\sigma \simeq 150~\kms$ for this initial nebular. 
Thus, the initial velocity of nebular is assumed in the form of $v_{\rm initial}=150~\kms \sin (\theta)^{1/2}$ rotated around the Z-axis, where $\theta$ is the polar angle with respect to Z-axis (see Fig. S8). 
The diffuse ISM is assumed with the number density of {$n_{\rm ism}= 10^3 (\cc) \times r_0/r(\rm pc)$ with $r_0= 1~\rm pc$ and } an initial velocity of 0. 
We acknowledge the spherical initial nebular might not be representative of the host galaxies of red quasars. 
It has been found from the rest-frame visible HST images that the majority of host galaxies of red quasars, including F2M0830, are merging systems \cite{Glikman2015, Urrutia2008}. 

In the next step, we propagate an isotropic outflow from the nuclear region { (as a ``nozzle'')} into the assumed environment. 
{The initial kinetic energy of wind is chosen based on the observed power of BAL outflow, which is found in general in the range of 1-10\% percent of the \ledd\ of quasars \cite{moe2009, borguet2013, chamberlain2015, he2019, Choi2020, He2022}. These studies motivate us to blow a high-velocity wind with a number density of $n_{\rm wind}=1.0\times 10^5~\cc$ and a velocity of $v_{\rm wind}=1\times 10^4~\kms$, which is launched at a galactocentric distance of 1pc and then propagates into the pre-setup ISM.  }
{We experiment with the initial kinetic energy to produce the resultant bubbles in line with the IFU data and found it to be $P_{\rm wind}$=$ 1.12\times 10^{46} \ergs$, corresponding to 8.6\% and 2.6\% of the assumed \lbol\ and \ledd. }
As shown in Fig. S8, the size of the outflow expands from about 10 kpc at 10 Myr to 20 kpc at 20 Myr. 
The expanding speed of the outflow surface (shock-wave) is $\sim$ 1000 $\kms$, consistent with our observational results. The number density of the bubble surface is 0.1-10$~\cc$.

Since the ionized outflow has been observed to be directional and appear in dual-ionized cones, we choose an opening angle of ionization core of $\Phi =45^{\circ}${\cite{he2018}} in our primary simulation run and $\Phi = 60^{\circ}$ as a comparison. 
The total power of wind within the ionization cone is $P_{\rm wind,~\Phi\leq45^{\circ}} = 0.3 P_{\rm wind}$. 
We note that only those outflows in the ionization cone are included in the following simulation (also see Fig. 4) since it was suggested that outflows are mostly observed within the ionization structure \cite{Zakamska2014}. 
We acknowledge that this assumption places the opening angle of the wind the same as that of the ionization cone. 
From the general model of quasar\cite{Zakamska2014}, the former is determined by circum-nuclear obscuring material, known as a torus. 
The observational evidence prefers a larger opening angle of wind than that of ionization cone \cite{Riffel2021}. 
However, in hydrodynamic simulations, the density distribution of the ISM is more important in terms of shaping the morphology of wind, rather than the opening angle of the wind \cite{Wagner2013}. 

%


The set equations of hydrodynamics for the interaction process are as follows:
\begin{equation}
 \frac{d\rho}{dt} + \rho \nabla \cdot \rm \mathbf{v} =0,
\label{eq1}
\end{equation}
\begin{equation}
\rho \frac{d\mathbf{v}}{dt} =- \nabla P - \rho\nabla \Phi,
\label{eq2}
\end{equation}
\begin{equation}
\frac{\partial e}{\partial t} +\nabla \cdot (e\mathbf{v} )=- P \nabla \cdot \mathbf{v},
\label{eq3}
\end{equation}
where $\rho$ is the density of the gas, $e$ is the internal energy density of the gas, $P=(\gamma -1)e$ is the gas pressure.
The viscosity and thermal conductivity are not included {in the simulation}.

{We also incorporate a radiation transfer receipt in the simulation. In detail, we include the cooling and heating terms following \cite{proga2000}: the Compton heating/cooling rate follows $G_{\rm Compton}=8.9\times 10^{-36}\xi_{X}(T_X-4T)$, the X-ray photoionization heating and recombination cooling rate follows $G_X=1.5\times 10^{-21}\xi_{X}^{1/4}T^{-1/2}(1-T/T_X)$, and the cooling function\cite{sutherland1993} for solar abundance is $L_{b,l}= 2.2\times 10^{-27}T^{0.5}+2.0\times 10^{-15}T^{-1.2} + 2.5 \times 10^{-24}$ for $T \geq 1 \times 10^5 K$, and $2.0 \times 10^{-31}T^{2.0}$ for $T < 1 \times 10^5 K$\cite{mou2017}, where $T_X$ is the temperature of X-ray radiation, which is 4 times the Compton temperature $T_c$. Considering $T_c$ is $\sim 1 \times 10^7 K$ for quasars\cite{yu2004}, we here set $T_X$ to be $4 \times 10^7 K$. The X-ray photoionization parameter is then defined as $\xi_X=L_X e^{-\tau_X(r)}/nr^2$, where $n$ is the gas density, $L_X$ is the intrinsic X-ray luminosity, and $\tau_X(r) = M_2 N_H\sigma_T$ is the X-ray optical depth in which the multiplier $M_2$ is 100 for $\xi_X < 10^5$ and 1 for $\xi_X \geq 10^5$. }

{Finally, we apply the photoionization simulation using \textsc{cloudy}\cite{Ferland2017} to generate the \oiii\ $\lambda$5007\AA\ emission coefficient as functions of ionization parameter ($U$) and gas density (see Fig. S9). The ionization parameter is defined as $U = Q_H / (4\pi r^{2} n_H c)$, where $Q_H$ is the source emission rate of hydrogen-ionizing photons, $r$ is the distance to the absorber from the source, $c$ is the speed of light, and $n_H$ is the hydrogen number density. We adopt the UV-soft Spectral Energy Distribution (SED) template \cite{dunn2010}, which is commonly used for high-luminosity radio-quiet quasars. We compute a set of models with the ionization parameters in a range of $-5 \leq \rm{log} U \leq 3$ with a step of $\Delta \rm{log}U = 0.1$, the gas density in a range of $-2 \leq \rm{log}(n_H) \leq 7$ with a step of $\Delta \rm{log}(n_H) =0.1$, and a solar metallicity $Z = Z\odot$. }
{For each grid, we obtain the \oiii\ $\lambda$5007\AA\ emission coefficient based on its ionization parameter and gas density from the bubble simulation. Note that the shielding effect due to any foreground cloud is included in the calculation of the ionization parameter. We then multiply the \oiii\ $\lambda$5007\AA\ emission coefficient by the length of the grid along the line of sight to obtain the surface brightness of a single grid. The final surface brightness is obtained by accumulating the surface brightness of all grids along the line of sight. }

To incorporate the viewing angle, we further obtain the 3D \oiii~emission coefficient map by rotating the 2D \oiii~emission coefficient map around the Y-axis and re-project the 3D \oiii~emission coefficient map with an inclination angle $i = 30^{\circ}$. This inclination angle is chosen to mimic the viewing angle of our red quasars. 
The results after having the simulation evolve for 12 Myr are shown in Fig. 4. 
In addition, we allow the simulated system to evolve for an additional period of 8 Myr. The corresponding \oiii\ surface brightness map is shown in the left panel of Fig. S7. 
As mentioned above, $\Phi = 60^{\circ}$ is adopted as a comparison to the effect of the opening angle of the ionization cone. The \oiii\ surface brightness map of $\Phi = 60^{\circ}$ is shown in the right panel of Fig. S7.

\smallskip
\noindent\textbf{Kinetic energy and mass flow of the galactic outflow from the simulation}

We integrate simulation grids to calculate the kinetic energy and mass flow of the galactic outflow.
The kinetic energy of the galactic outflow is calculated as follows:
\begin{equation}
{\dot{E}_{\rm out}}=\frac{1/2 \int_{0.5kpc}^{40kpc} \int_{0}^{\pi} \int_{0}^{2\pi} n_{\rm e}v^3m_{\rm p}r^2\sin (\theta)drd\theta d\psi}{T},
\label{eq4}
\end{equation}
where $n_{\rm e}$ is the number density of gas, $m_{\rm p}$ is the mass of proton {and $T$ is the total evolution time scale in the simulation}. 
{We note that we assume the gas is singly ionized, i.e., contains only ionized hydrogen. We do not take into account the doubly ionized Helium, which would introduce a small correction of $\sim$1.09 \cite{Deharveng2000}. }
As shown in the left panel of Fig. S10, the kinetic energy of galactic outflow driven by the central quasar winds is ${\dot{E}_{\rm out}} =3.16\times 10^{43}\ergs$, about 3\% of power of quasar winds. 
The kinetic energy of the bubble shell (where the \oiii\ emission coefficient exceeds $10^{-24}~\ergs \cc $) is $\sim$10\% of that of the outflow.

The mass flow rate of the galactic outflow is calculated as follows:
\begin{equation}
{\dot {M}_{\rm out}}=\frac{\int_{0.5kpc}^{40kpc} \int_{0}^{\pi} \int_{0}^{2\pi} n_{\rm e}m_{\rm p}r^2\sin (\theta)drd\theta d\psi}{T}. 
\label{eq5}
\end{equation}
The mass flow rate of quasar wind and outflow as functions of evolved time are shown in the right panel of Fig. S10. 
The mass flow rate of input quasar winds is ${\dot{M}_{\rm wind}}\sim$100 \mpyr. 
The mass flow rate of outflow is ${\dot{M}_{\rm out}} \sim$ 1000 \mpyr, one order of magnitude higher than that of input quasar winds. 
The energetic mass outflow from simulation tentatively suggests that the quasar winds might have sufficient interaction with the interstellar medium and might be capable of shaping the evolution of its host galaxy \cite{he2019, He2022}.
Nevertheless, we acknowledge that the predicted ${\dot{M}_{\rm out}}$ might be an upper limit since it is critically determined by the shape of the assumed galaxy. A lower ${\dot{M}_{\rm out}}$ might be expected for a disk galaxy with its wind breaking out into a pair of bubbles perpendicular to the disk \cite{Wagner2013}.





\section*{Acknowledgments}



\textbf{Acknowledgments}: We appreciate the informative discussion with Feng Yuan. This project is based on the data obtained with Gemini telescope (programme ID: GN-2014A-Q-19, PI: G. Liu). We thank the scientists and telescope operators at Gemini telescope for their help.
This project used GEMINI package the Image Reduction and Analysis Facility (IRAF) that is distributed by the National Optical Astronomy Observatories which is operated by the Association of Universities for Research in Astronomy, Inc. under cooperative agreement with the National Science Foundation.\\

\textbf{Funding}: This work was supported by the research grants from the China Manned Space Project (the 2nd-stage CSST science project: {\em Investigation of small-scale structures in galaxies and forecasting of observations}, No. CMS-CSST-2021-A06 and CMS-CSST-2021-A07), the National Natural Science Foundation of China (No. 12273036, 11421303), the Fundamental Research Funds for the Central Universities (No. WK3440000005), the support from Cyrus Chun Ying Tang Foundations, and the lateral fund from Shanghai Astronomical Observatory (No. EF2030220007).
L.S. acknowledges the National Natural Science Foundation of China (No. 12003030). Z.H. is supported by National Natural Science Foundation of China (No. 12222304, 12192220 and 12192221).
E.G. acknowledges the generous support of the Cottrell Scholar Award through the Research Corporation for Science Advancement. E.G. is grateful to the Mittelman Family Foundation for their generous support. 
G.M. is supported by National Natural Science Foundation of China (No. 11833007). \\
\textbf{Author Contributions}: 
L.S. reduced the data, performed the scientific analysis, and led the writing of the manuscript. G.L. performed the data acquisition, early-stage data reduction and analysis, and co-led the manuscript writing. G.L., N.Z., J.G. and E.G. conceived, designed, and initiated the project. G.L. and N.Z. conducted the observations and co-led the scientific analysis and interpretation. Z.H. and N.Z. developed the theoretical aspect of this work. Z.H. conducted numerical simulations, prepared the related content of the manuscript, and co-led the scientific interpretation. 
G.M. wrote the code for hydrodynamic simulations based on the open source code {\sc zeusmp}, and contributed part of the simulation-related text. 
All authors discussed and commented on the content of the paper. 
\\
\textbf{Competing interests}: The authors declare that they have no competing interests. \\
\textbf{Data and materials availability}: All data needed to evaluate the conclusions in the paper are presented in the paper and/or the Supplementary Materials.  

\clearpage

\begin{figure*} 
\includegraphics[width=\textwidth]{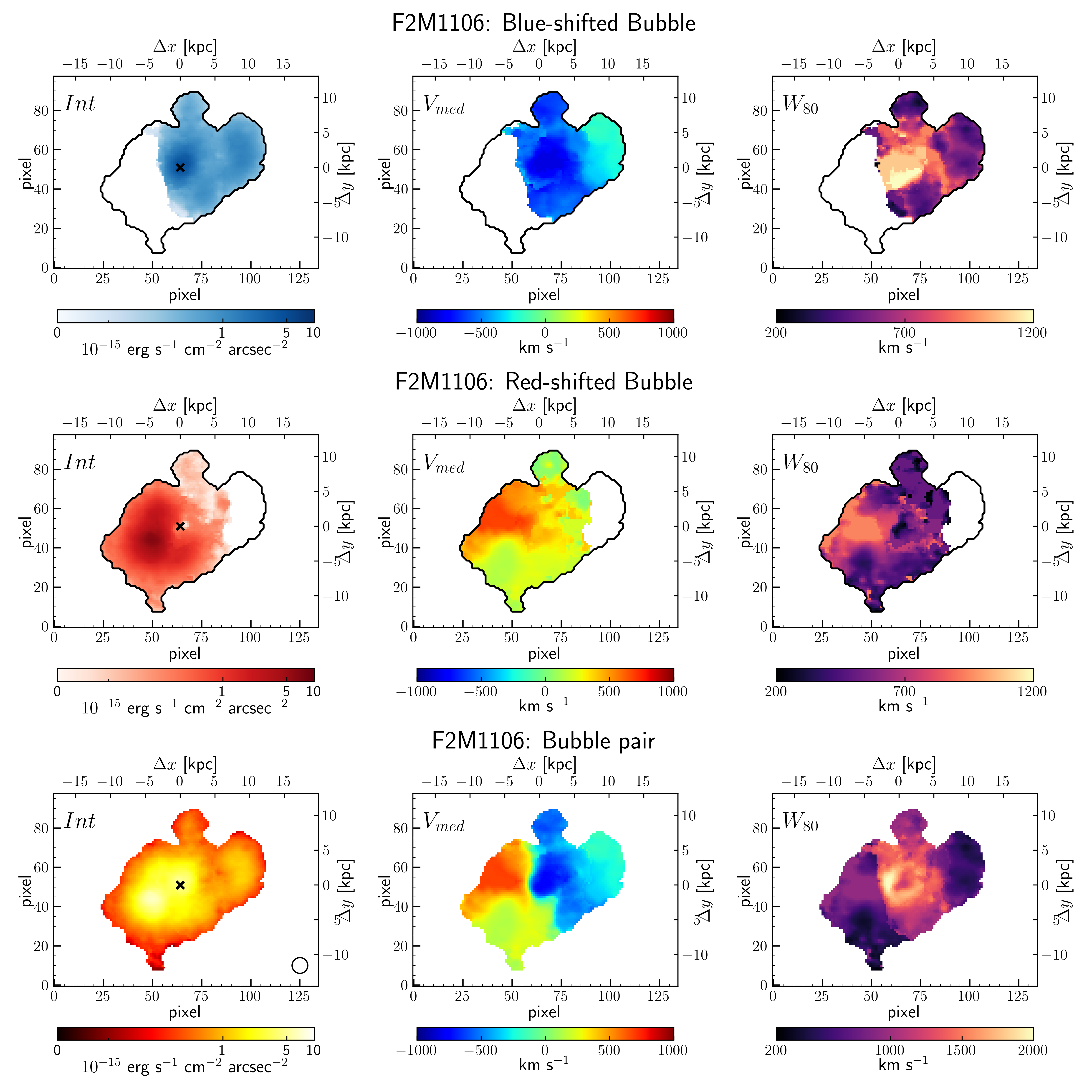}
\caption{{{\bf Fig. 1. The maps of ionized gas nebulae for F2M1106.} From left to right panels are the surface brightness integrated luminosity ($Int$), the line-of-sight velocity ($V_{\rm med}$), and velocity dispersion ($W_{80}$) maps of the \oiii\ $\lambda$5007 \AA\ line.} From top to bottom panels are the maps of blue-shifted and red-shifted parts of the bubbles and together. The intensity maps are shown on a logarithmic scale in units of 10$^{-15}$ \ergscmarc. The $V_{\rm med}$ and $W_{80}$ maps are shown in units of km s$^{-1}$. In each spaxel, the \oiii\ flux, $V_{\rm med}$, and $W_{80}$ are calculated from a multi-Gaussian fit to the spectral profile of the line. Only spaxels where the peak of the \oiii\ line is detected with S/N $>$ 5 are plotted. The PSF is subtracted. {The black crosses in the \textit{left} panels indicate the peak-flux location of the PSF for this red quasar (see Methods). The circle in the \textit{bottom left} panel indicates the seeing (see Table S1). North is up, and east is left. } }
\label{fig:maps1}
\end{figure*}


\begin{figure*} 
\includegraphics[width=\textwidth]{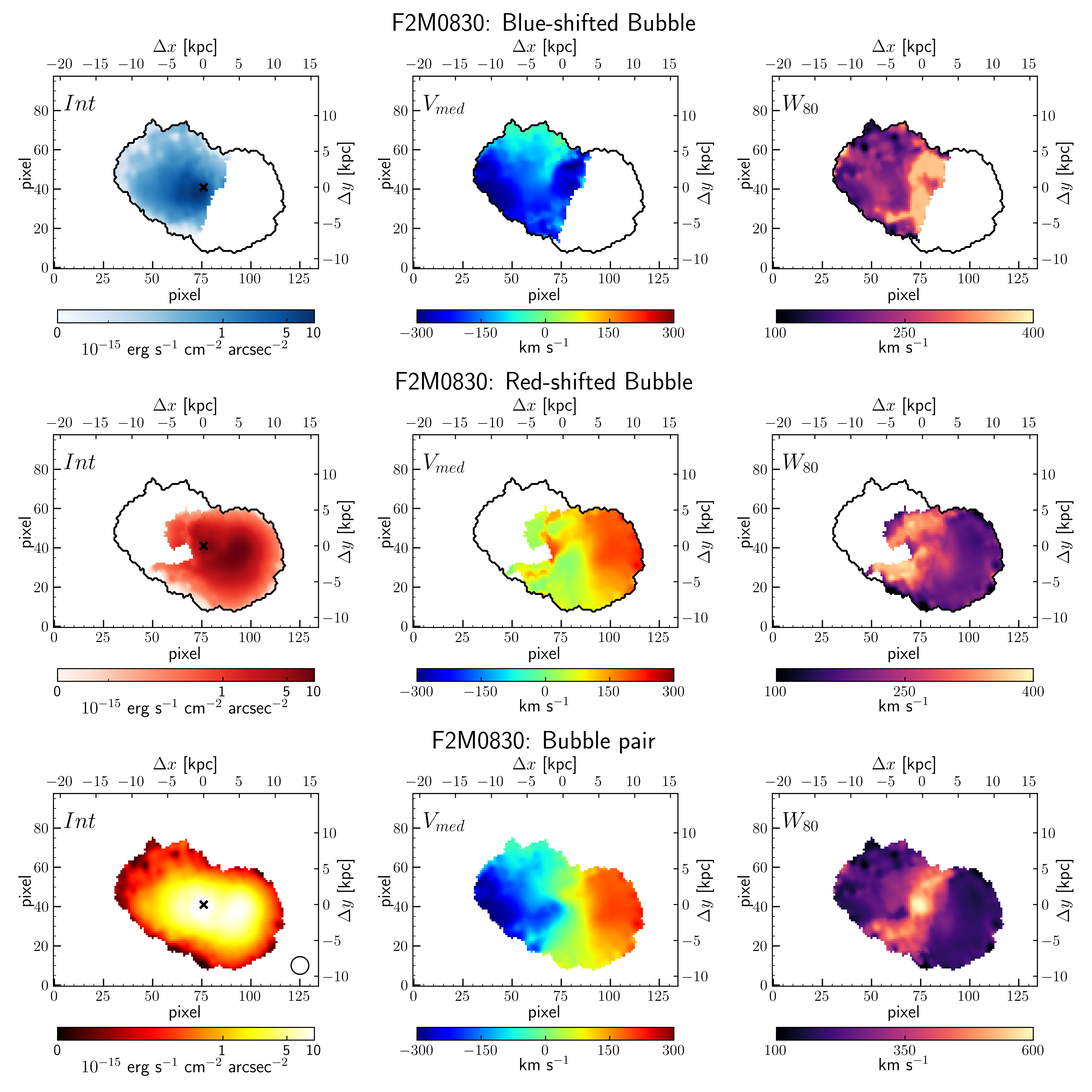}
\caption{{{\bf Fig. 2. The maps of ionized gas nebulae for F2M0830.} Same as Fig. 1. }}
\label{fig:maps2}
\end{figure*}

\begin{figure*} 
\includegraphics[width=\textwidth]{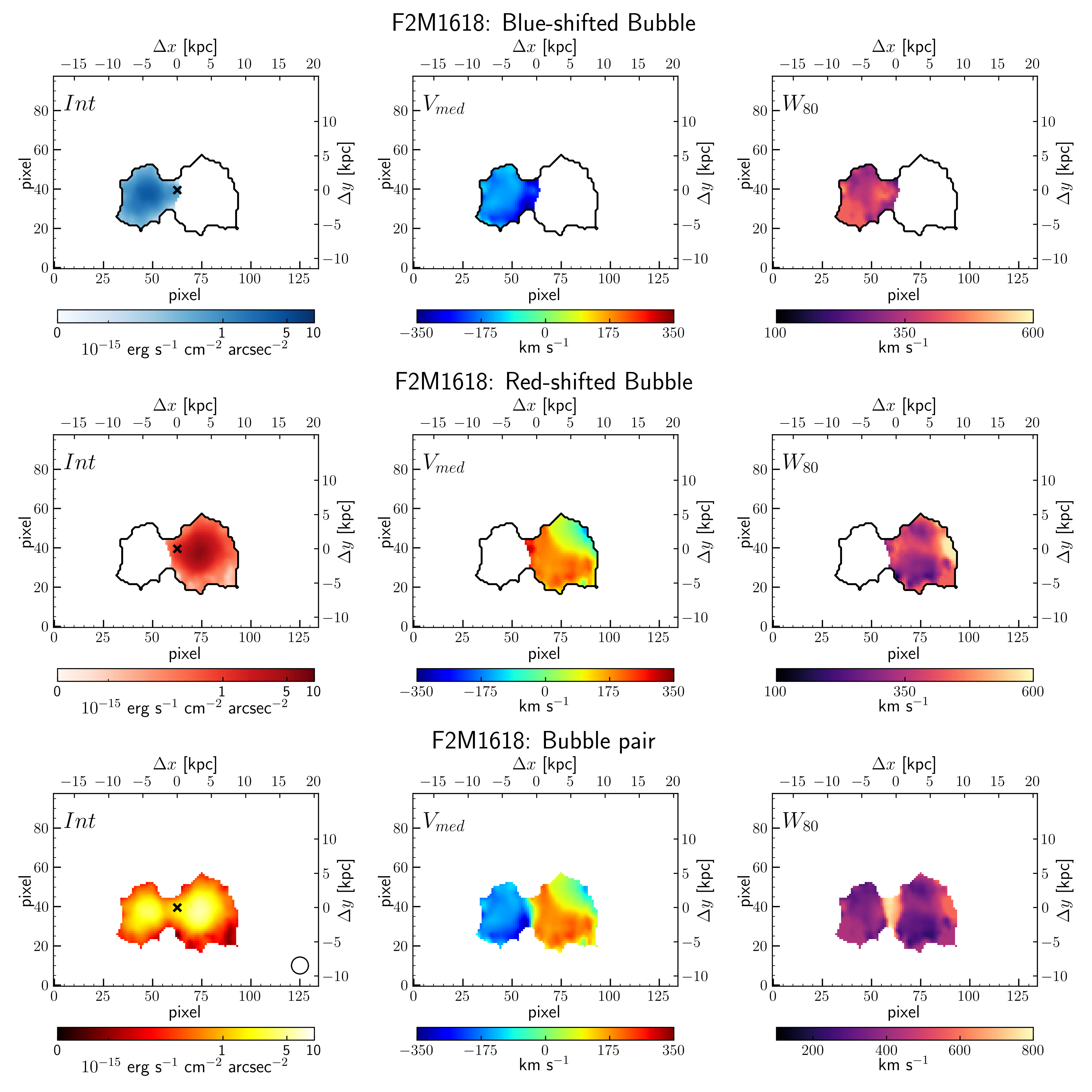}
\caption{{{\bf Fig. 3. The maps of ionized gas nebulae for F2M1618.} Same as Fig. 1. } }
\label{fig:maps3}
\end{figure*}

\begin{figure*}
\centering
\includegraphics[width=\textwidth]{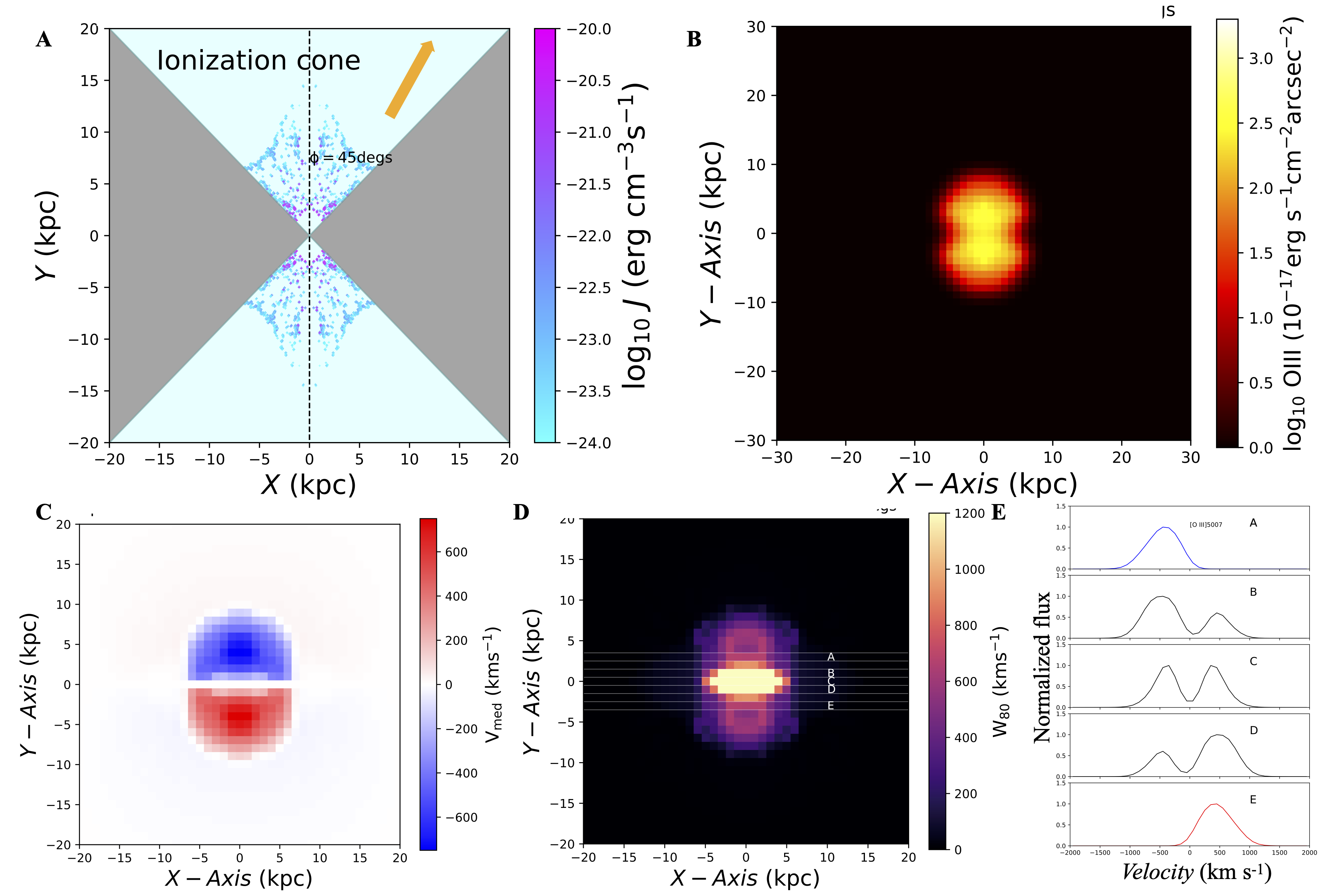}
\caption{\textbf{Fig. 4. The results of our two-dimensional (2D) hydrodynamics simulation, well reproduced the morphology and kinematics of the observed superbubbles.} Panels are (\textit{A}) the \oiii~emission coefficient map, (\textit{B}) the \oiii\ surface brightness map, (\textit{C}) the median velocity map, (\textit{D}) the $W_{80}$ map, and (\textit{E}) the \oiii~emission line profiles at the location marked as white lines in the $W_{80}$ map (panel \textit{D}). }
\label{fig:bubble}
\end{figure*}

\begin{figure*}
\centering
\includegraphics[width=0.5\textwidth]{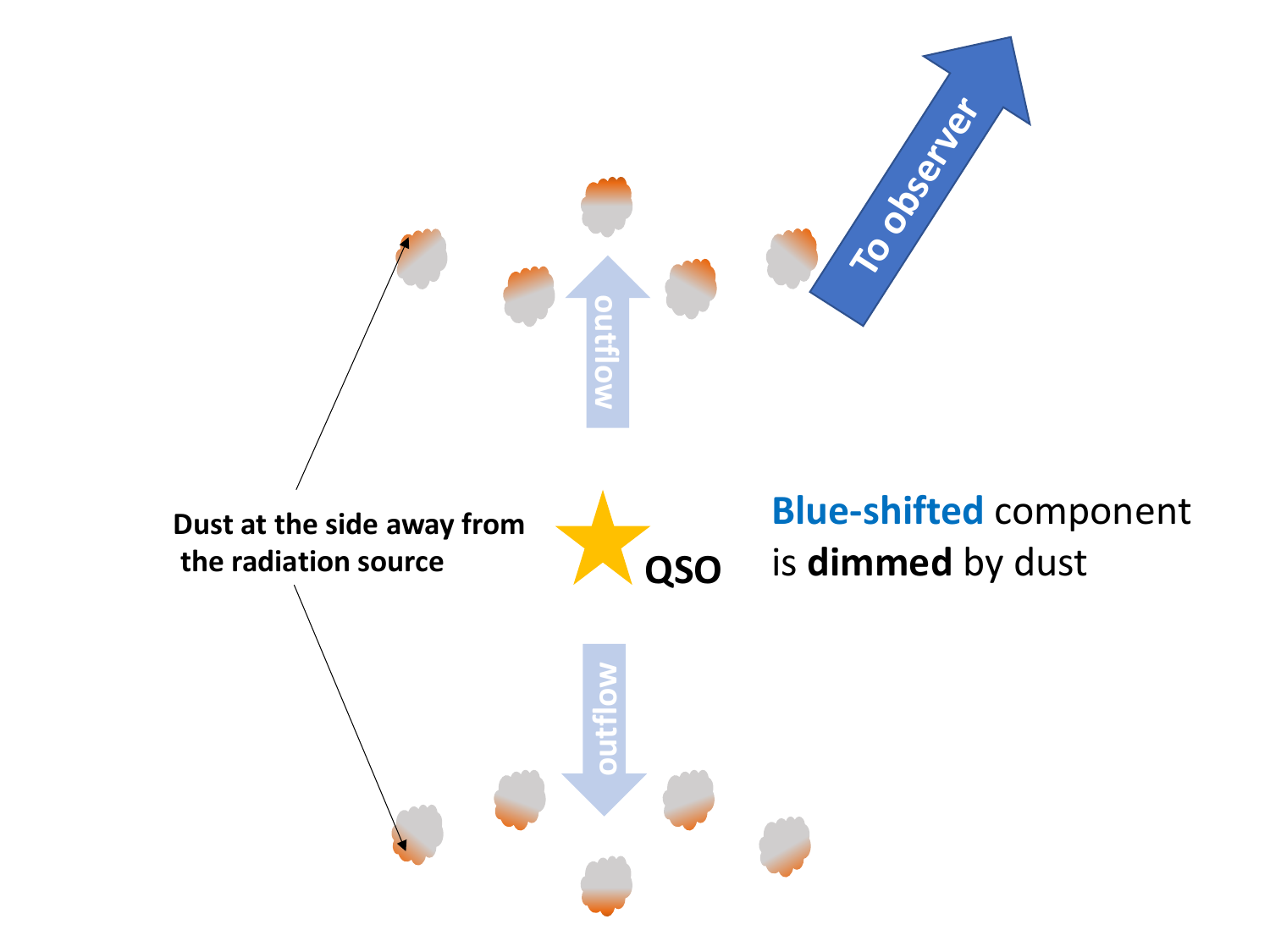}
\caption{\textbf{Fig. 5. The illustration of brighter red-shifted bubbles and fainter blue-shifted bubbles in our observation.} The dust exists predominantly on the side of the outflow gas away from the radiation source, so that the blue-shifted component will be dimmed by the dust. }
\label{fig:dust}
\end{figure*}

\begin{table*}[htp]
\caption{\bf Table 1. Luminosity, morphology and kinematic properties of the [O \textsc{iii}] outflow of the three red quasars}
\begin{center}
\begin{threeparttable}
\footnotesize
\begin{tabular}{c ccc cc ccccc}
\hline 
\hline 
Object & $L_{\rm [O\;{\scriptscriptstyle III}]}$ & $L_{\rm [O\;{\scriptscriptstyle III}]}^{red}$ & $L_{\rm [O\;{\scriptscriptstyle III}]}^{blue}$ & $R_\mathrm{{5\sigma}}$ & $\epsilon_\mathrm{{5\sigma}}$ & $R_\mathrm{{int}}$ & $\epsilon_\mathrm{{int}}$ &  $\Delta v_{max}$ & $\langle W_{80} \rangle$ & $W_{80, \rm max}$ \\
 & log(erg/s) & log(erg/s) & log(erg/s) & kpc & & kpc & & km/s & km/s & km/s\\
 (1) & (2) & (3) & (4) & (5) & (6) & (7) & (8) & (9)  & (10) & (11)\\
\hline 
F2M0830 & 43.09 & 42.90 & 42.65 & {12.3} & {0.76} & {10.7} & {0.77} & 394 & 222 & 480 \\
F2M1106 & 42.87 & 42.64 & 42.50 & {11.8} & {0.73} & {11.7} & {0.74} & 1216 & 931 & 1602 \\
F2M1618 & 42.56 & 42.37 & 42.12 & {8.7} & {0.82} & {8.3} & {0.81} & 671 & 397 & 703 \\
\hline 
\end{tabular}
\begin{tablenotes}
\item Notes: (1) Object name. (2) Total luminosity of the [O \textsc{iii}] $\lambda$5007 \AA\ line. (3, 4) [O \textsc{iii}] luminosity of the red- and blue-shifted bubbles. (5, 6) Semi-major axis (in kpc) and ellipticity of the best-fitting ellipse which encloses pixels with S/N $\geq$ 5 in the [O \textsc{iii}] $\lambda$5007 \AA\ line map. (7, 8) Isophotal radius (semi-major axis) and ellipticity at the intrinsic limiting surface brightness (corrected for cosmological dimming) of $10^{-15}/(1+z)^4$ erg s$^{-1}$ cm$^{-2}$ arcsec$^{-2}$. (9) Maximum difference in the median velocity map. (10, 11) Median and maximum $W_{80}$ values in the $W_{80}$ map. For columns 7-9, only spaxels with S/N $\geq$ 5 are used and the 5 per cent tails on either side of the distribution are excluded to minimize the effect of the noise. 
\end{tablenotes}
\end{threeparttable}
\end{center}
\label{tab:properties}
\end{table*}


\newpage

\section*{Supplementary materials}
Figs. S1 to {S10} \\
Tables S1 to S2\\



\clearpage

\begin{figure*} 
\includegraphics[width=\textwidth]{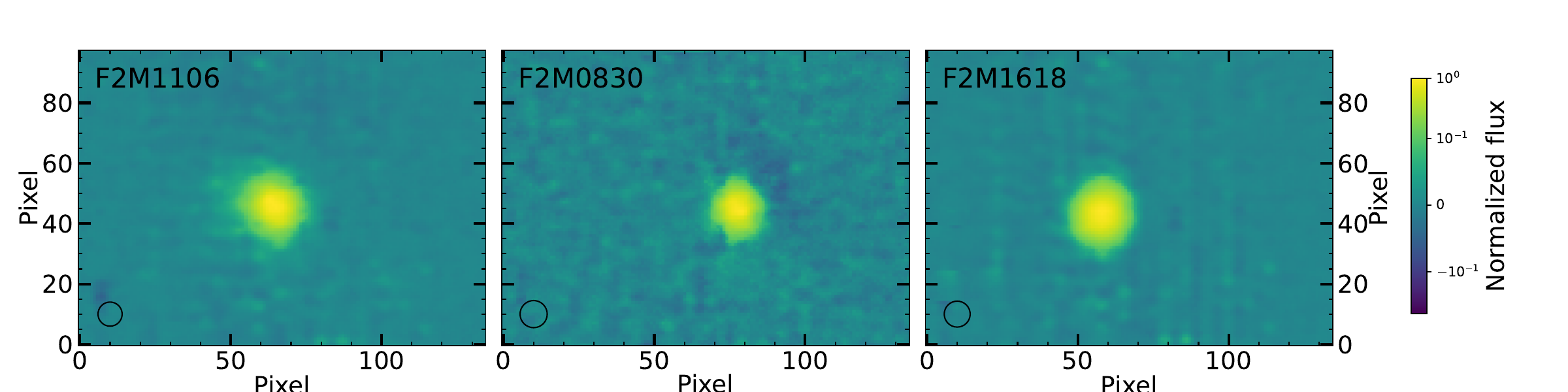}
\caption{\textbf{Fig. S1. The PSF of our red quasars. } For F2M1618, the PSF in the vicinity of [O \textsc{iii}] is interpolated between the normalized median images in $\lambda_{rest} =$ 4970 - 4980 \AA\ and $\lambda_{rest} =$ 5030 - 5050 \AA. While, due to the wavelength coverage of IFU data for the other two red quasars (F2M0830 and F2M1518), their PSF are the normalized median images in $\lambda_{rest} =$ 5030 - 5050 \AA. The seeing at the observing site is depicted by the open circle. }
\label{fig:psf}
\end{figure*}

\begin{figure*} 
\includegraphics[width=\textwidth]{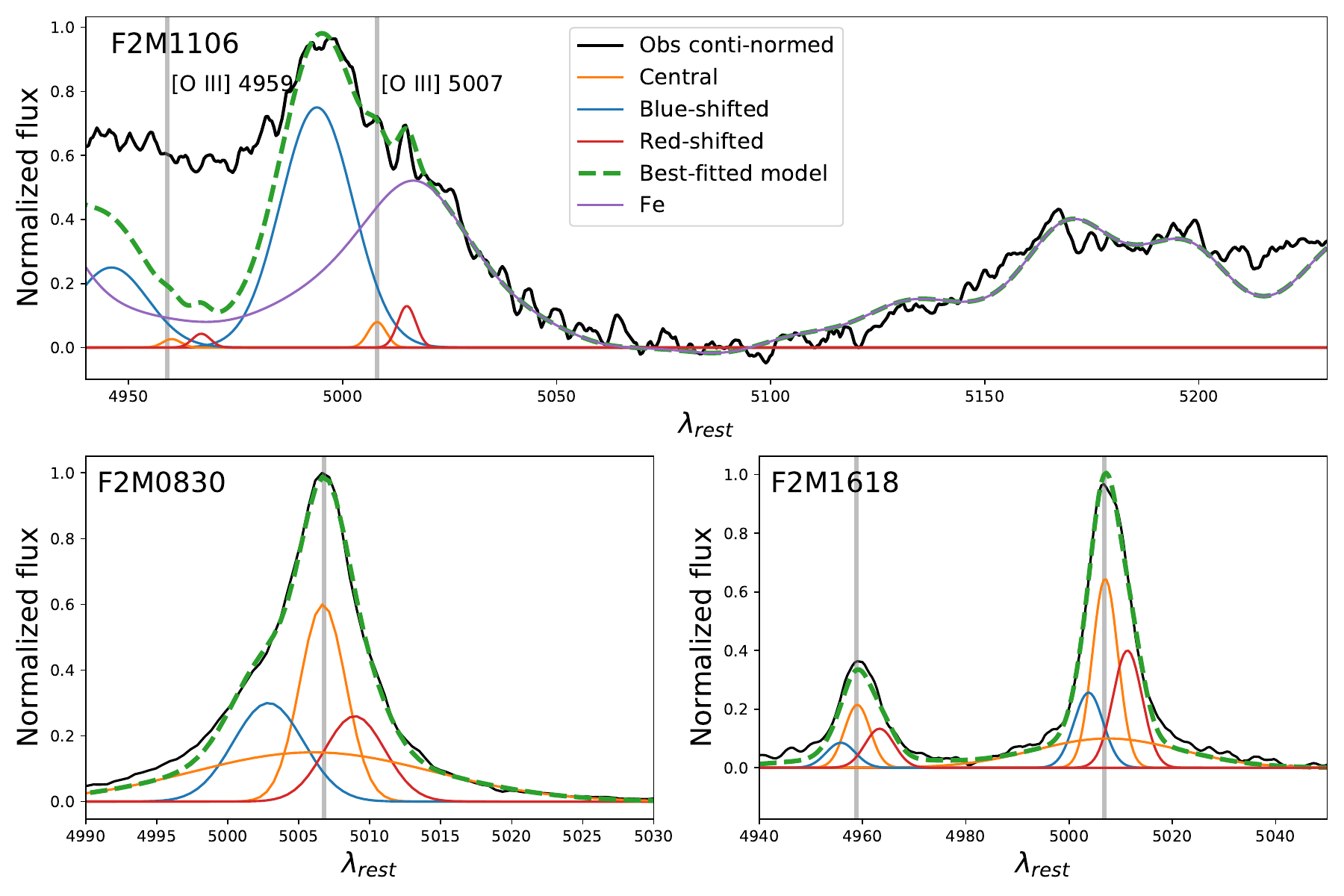}
\caption{\textbf{Fig. S2. The quasar spectra of our red quasars. } For each quasar, we generate its central spectrum by combining the central 5$\times$5 spaxels and normalizing in $\lambda_{rest} =$ 5060 - 5100 \AA. Spectra are smoothed with a box filter of 5 pixels for visualization purposes. The modeled spectra consist of three/four Gaussians (solid orange line for the quasar dominated, solid blue line for blue-shifted and solid red for red-shifted components) and Fe \textsc{ii} emission (purple) if detected. The total reconstructed spectra are shown with a dashed green line. The quasar
spectrum is constructed by removing the recovered red- and blue-shifted \oiii\ components in the
central spectra. }
\label{fig:central_spec}
\end{figure*}

\begin{figure*} 
\centering
\includegraphics[width=0.5\textwidth]{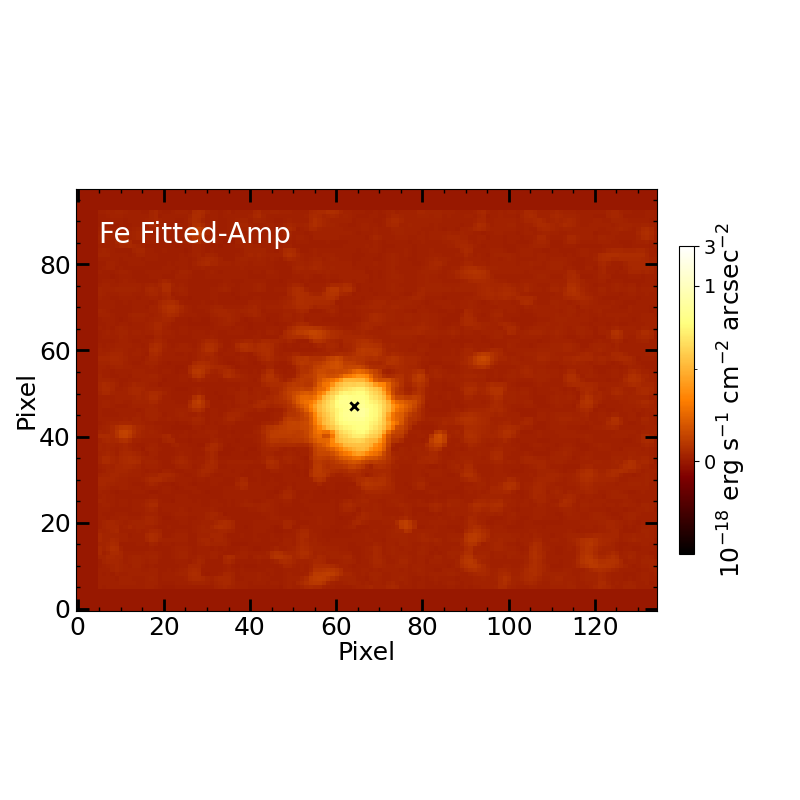}
\vspace{-10mm}
\caption{\textbf{Fig. S3. The intensity map of F2M1106 Fe \textsc{ii} emission.} For each spaxel, the intensity of Fe \textsc{ii} emission is determined by fitting to the continua in rest-frame $\lambda$ 5100 - 5250 \AA\ using the Fe \textsc{ii} template\cite{Boroson1992} and smoothed using a Gaussian kernel whose width is one of the free fitting parameters. The intensity map of Fe \textsc{ii} emission reveals a point source with the same size as the PSF and centered at the central spaxel. }
\label{fig:Fe_map}
\end{figure*}

\begin{figure*} 
\centering
\includegraphics[width=1\textwidth]{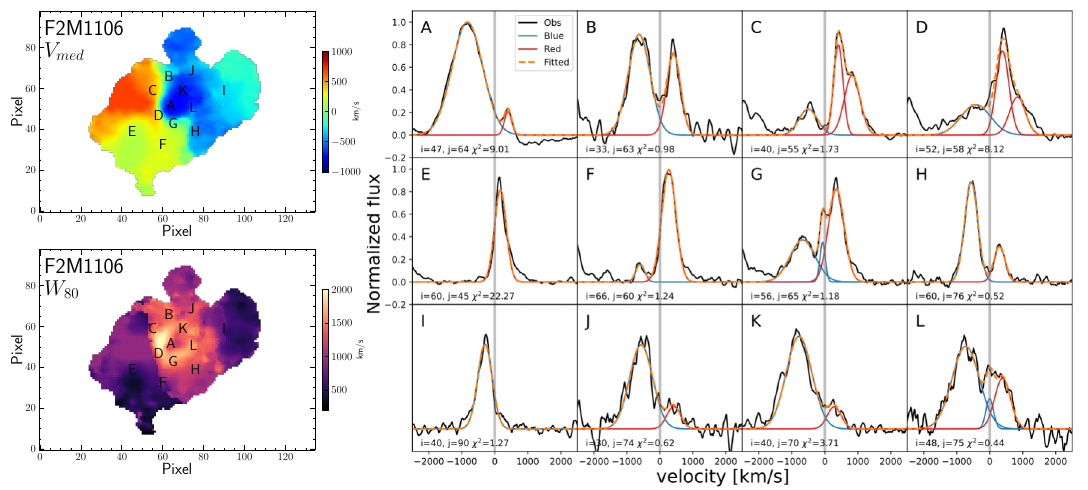}
\vspace{-5mm}
\caption{\textbf{Fig. S4. Example spectra in the vicinity of the \oiii\ $\lambda$5007 \AA\ line in spaxels for F2M1106. } \textit{Left:} The $V_{\rm med}$ (\textit{left top}) and $W_{80}$ (\textit{left bottom}) maps the same as in Fig 1. \textit{Right:} Example spectra in the vicinity of the  \oiii\ $\lambda$5007 \AA\ line in spaxels, fitted by multi-Gaussian components. Each spectrum is smoothed by a 5 \AA\ box filter shown in black. The best-fitted distribution is marked in orange with each Gaussian shown in blue and red for the blue- and red-shifted components. The location of each spectrum is marked in the left panels. }
\label{fig:maps_gaussians_F2M1106}
\end{figure*}

\begin{figure*} 
\centering
\includegraphics[width=\textwidth]{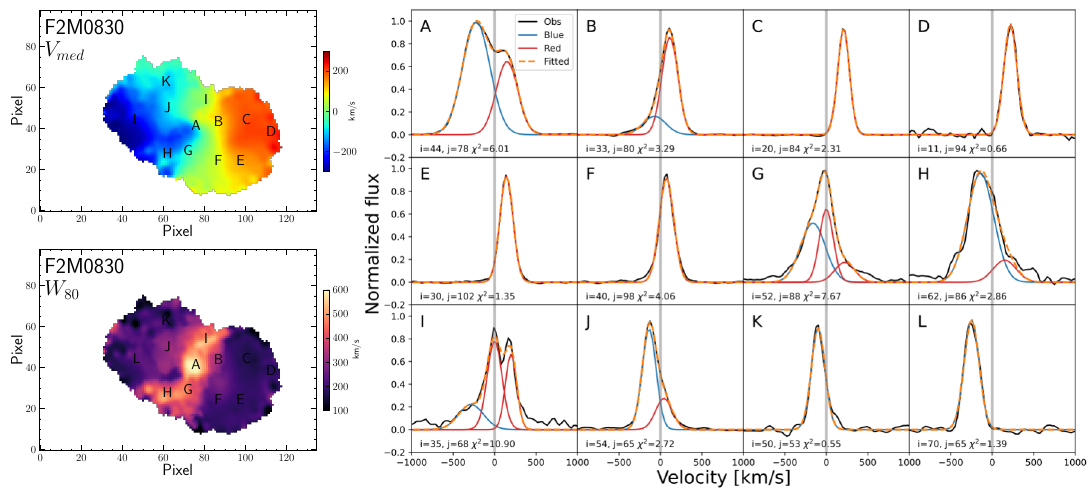}
\vspace{-5mm}
\caption{\textbf{Fig. S5. Example spectra in the vicinity of the \oiii\ $\lambda$5007 \AA\ line in spaxels for F2M0830. } }
\label{fig:maps_gaussians_F2M0830}
\end{figure*}

\begin{figure*} 
\centering
\includegraphics[width=\textwidth]{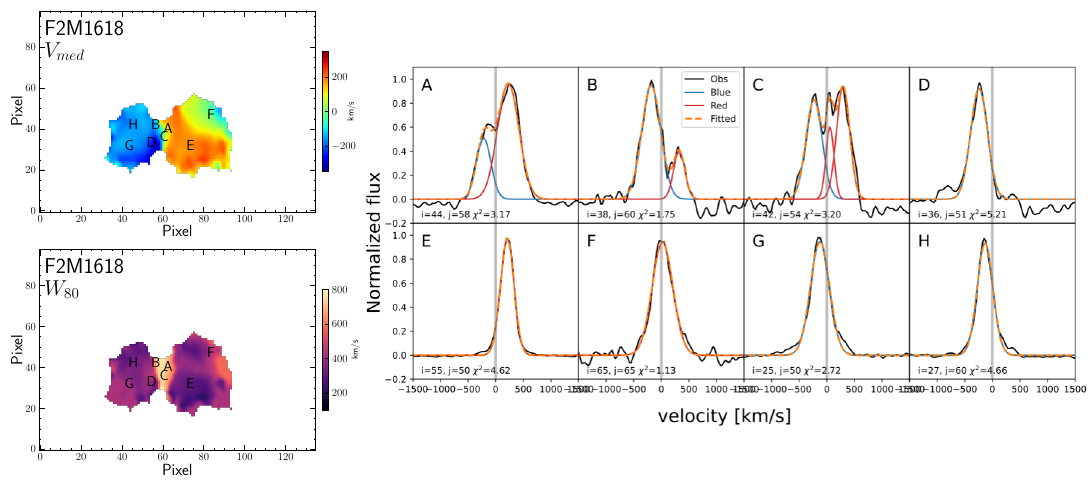}
\vspace{-5mm}
\caption{\textbf{Fig. S6. Example spectra in the vicinity of the \oiii\ $\lambda$5007 \AA\ line in spaxels for F2M1618. } }
\label{fig:maps_gaussians_F2M1618}
\end{figure*}

\begin{figure*}
\centering
\includegraphics[width=\textwidth]{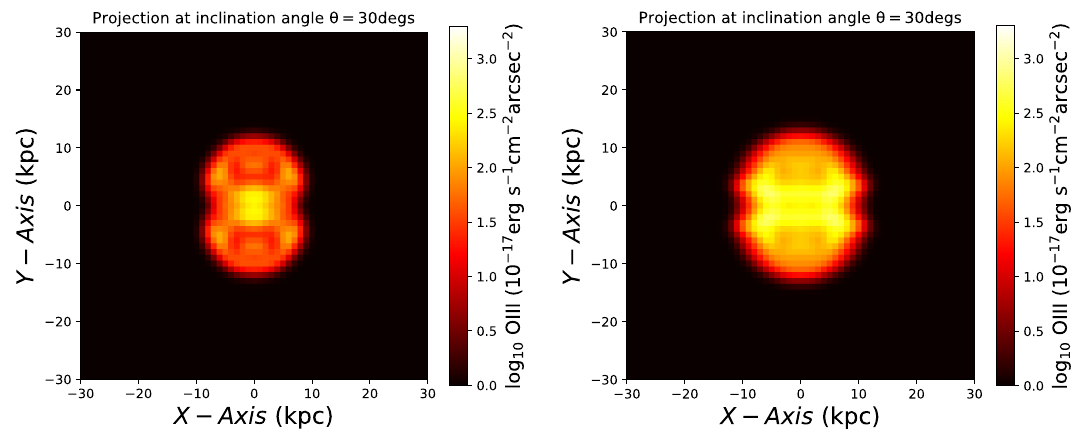}
\caption{\textbf{Fig. S7. The projected \oiii\ surface brightness maps of a simulated outflow evolved for 20 Myr with half opening angles of ionization cone being 45$^{\circ}$ (\textit{left}) and 60$^{\circ}$ (\textit{right})}. The same inclination angle $i$ =30$^{\circ}$ is used as in Fig. 4. 
The simulated nebulae gradually grow into a sphere. In addition, a spherical shape could be obtained by increasing the opening angle of the ionization cone.}
\label{fig:bubble2}
\end{figure*}

\begin{figure*}
\centering
\includegraphics[width=\textwidth]{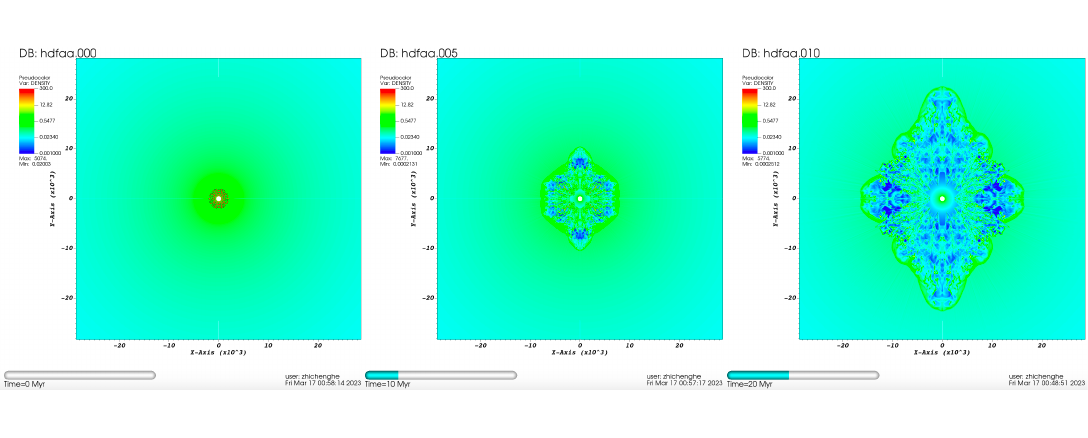}
\caption{\textbf{Fig. S8. The density distribution of ISM driven by quasar winds in the simulation} From left to right panels shows the density distribution of nebula at 0, 10, and 20 Myr. }
\label{fig:density}
\end{figure*}

\begin{figure*}
\centering
\includegraphics[width=0.4\textwidth]{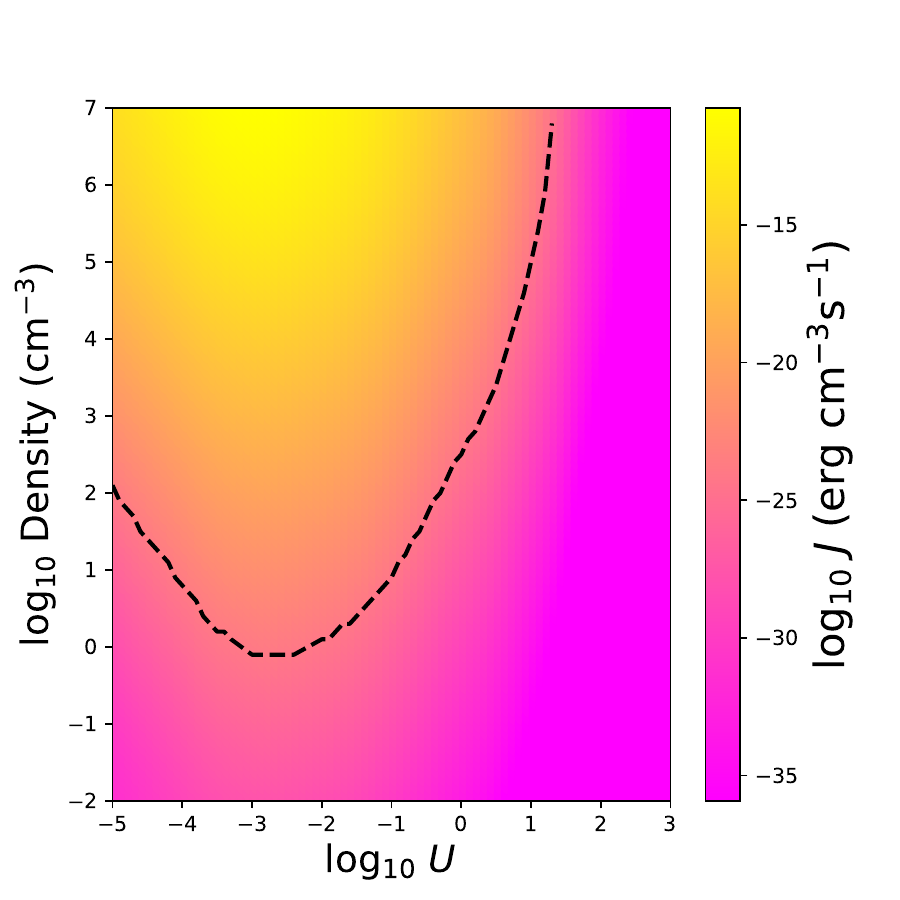}
\caption{\textbf{Fig. S9. \oiii\ emission coefficient $J$ at different  gas density and ionization parameters calculated by the CLOUDY. } The
dashed black curve marks the boundary J=$10^{-24} \ergs \cc$. }
\label{fig:coeff}
\end{figure*}

\begin{figure*}
\centering
\includegraphics[width=\textwidth]{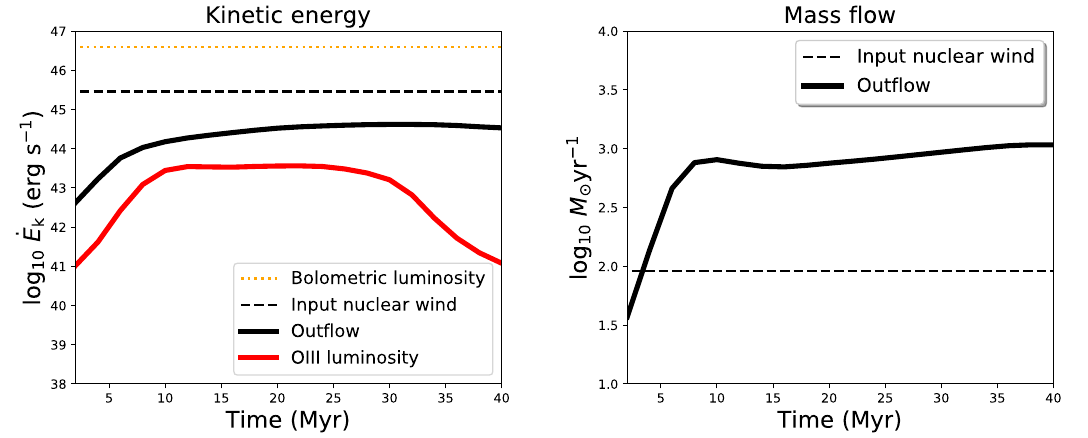}
\caption{\textbf{Fig. S10. Kinetic energy and mass flow of the galactic outflow.}
\textit{Left}: the bolometric luminosity and the power of input central quasar winds in the ionization cone are $\lbol =3.9\times 10^{46}\ergs$ and $P_{\rm wind}$=8.6\% \lbol=$ 3.35\times 10^{45}\ergs$, respectively. The kinetic energy of galactic outflow driven by the central quasar winds is ${\dot {E_{\rm out}}} = 3.0\times 10^{44}\ergs$, $\sim$10\% of the input central quasar winds. 
\textit{Right}: the mass flow rates of input central quasar winds in the ionization cone and galactic outflow are ${\dot {M}_{\rm wind}}\sim$100\mpyr and ${\dot{M}_{\rm out}}\sim$1000\mpyr. The outflow is one order of magnitude higher than the input central quasar winds.}
\label{fig:energy}
\end{figure*}

\clearpage

\begin{table*}
\caption{\textbf{Table. S1. Observation properties of our red quasar sample. }}
\begin{center}
\begin{threeparttable}
\footnotesize
\begin{tabular}{ccccccccc}
\hline 
\hline 
Object & $\alpha$ & $\delta$ & $z$ & P.A. & $t_{\rm exp}$ & Seeing & $R^{\rm PSF}_{\rm eff}$ & $\sigma$\\
& & & & deg & s & arcsec & arcsec & erg s$^{-1}$ cm$^{-2}$ arcsec$^{-2}$ \\
(1) & (2) & (3) & (4) & (5) & (6) & (7) & (8) & (9) \\
\hline 
F2M0830 & 08:30:11.13 & +37:59:51.7 & 0.414 & 69  & 1800$\times$2 & 0.45 & 0.57 & 4.3e-18\\
F2M1106 & 11:06:48.32 & +48:07:12.3 & 0.435 & 0   & 1800$\times$2 & 0.40 & 0.51 & 5.9e-18\\
F2M1618 & 16:18:09.74 & +35:02:08.8 & 0.447 & 270 & 1800$\times$2 & 0.43 & 0.59 & 4.6e-18\\
\hline 
\end{tabular}
\begin{tablenotes}
\item Notes: (1) Object name. (2-4) Coordinates and redshift. (5) Position angle of the field of view (degrees). (6) Exposure time (seconds) and the number of exposures. (7) Seeing at the observing site determined by averaging the directly measured FWHM of a sample of field stars in the acquisition image taken before the science exposure. (8) The effective radius of the PSF. (9) The surface brightness sensitivity of our \oiii\ maps, determined from each spaxel in the wavelength range of $\lambda_{\rm rest}$ = 5050 - 5080 \AA. 
\end{tablenotes}
\end{threeparttable}
\end{center}
\label{tab:obs}
\end{table*}

\begin{table*}
\caption{\textbf{Table. S2. Properties of the red quasar sample.}}
\begin{center}
\begin{threeparttable}
\footnotesize
\begin{tabular}{l c c c c c c c c c c }
\hline 
\hline 
Object & E(B-V) & $L_{bol, 5100, corr}$ & $\nu L_{\nu, 6\mu m}$  & $\nu L_{\nu, 12\mu m}$ & $\nu L_{bol, 12\mu m}$ & $f_{1.4GHz}$ & $f_{20cm}$ & $R$ \\
 &  &   log(erg/s) &  log(erg/s) & log(erg/s) & log(erg/s) & mJy &  mJy & \\
 (1) & (2) & (3) & (4) & (5) & (6) & (7) & (8) & (9) \\
\hline 
F2M0830 & 0.74 & 45.99 & 45.38 & 45.54 & 46.50 & 6.42  & 6.79  & -4.68 (-4.66) \\
F2M1106 & 0.44 & 46.24 & 45.87 & 45.86 & 46.81 & 10.03 & 10.50 & -4.93 (-4.91)\\
F2M1618 & 0.69 & 46.09 & 45.72 & 45.68 & 46.63 & 14.96 & 34.25 & -4.58 (-4.23) \\
\hline 
\end{tabular}
\begin{tablenotes}
\item Notes: (2) the extinction E(B-V) are adopted from ref.\cite{Glikman2012} using the full continuum approach. (3) Bolometric luminosity derived from rest-frame 5100\AA\ luminosity with a bolometric correction of BC$_{5100}$ = 9.65 \cite{Shen2011, Richards2006} and additional dust correction to $L_{5100}$ with $A_{V}$ = 3.1 E(B-V) (4) \& (5) Luminosity at rest-frame 6~$\mu$m and 12~$\mu$m interpolated from WISE photometry. (6) Bolometric luminosity derived from rest-frame 12~$\mu$m luminosity with a bolometric correction of 9 from ref. \cite{Richards2006}. (7) Peak flux densities at 1.4 GHz obtained from FIRST \cite{Becker1995} in B configuration ($\sim5^{\prime\prime}$ beam) (8) Peak flux densities at 20~cm from ref. \cite{Glikman2007} in CnD configuration ($\sim15^{\prime\prime}$ beam) for F2M0830 and F2M1618 and NVSS \cite{Condon1998} in D configuration (45$^{\prime\prime}$ beam) for F2M1106 (9) Radio loudness using peak flux densities at 1.4~GHz from col. 7 and using flux at 20~cm from col. 8 shown in the parentheses.  
\end{tablenotes}
\end{threeparttable}
\end{center}
\label{tab:general}
\end{table*}



\end{document}